\documentclass[journal, onecolumn]{IEEEtran}
\usepackage{amsmath,amsfonts,subfigure,adjustbox}
\usepackage{algorithmic}
\usepackage{algorithm}
\usepackage{array}
\usepackage[caption=false,font=normalsize,labelfont=sf,textfont=sf]{subfig}
\usepackage{textcomp}
\usepackage{stfloats}
\usepackage{url}
\usepackage{verbatim}
\usepackage{graphicx}
\usepackage{cite}
\usepackage{makecell}
\usepackage{color}
\newtheorem{Remark}{Remark}
\newtheorem{Definition}{Definition}
\newtheorem{Theorem}{Theorem}
\newtheorem{Lemma}{Lemma}

\begin{document}
	
	%\title{Low oversampling ambiguity zone sequence design for channel estimation in doubly selective channel}
	%\title{Low Oversampled Ambiguity Zone Sequences and its Application in Channel Estimation in Doubly Selective Channels}
	%	
	%	\author{IEEE Publication Technology,~\IEEEmembership{Staff,~IEEE,}
		\title{Oversampled Low Ambiguity Zone Sequences for Channel Estimation over Doubly Selective Channels}
		
		\author{Zhi Gu, Zhengchun Zhou, Pingzhi Fan, Avik Ranjan Adhikary, and Zilong Liu
}
		
		\markboth{Journal of \LaTeX\ Class Files,~Vol.~14, No.~8, August~2021}%
		{Shell \MakeLowercase{\textit{et al.}}: A Sample Article Using IEEEtran.cls for IEEE Journals}
		
		
		\maketitle
		
		\begin{abstract}
			Pilot sequence design over doubly selective channels (DSC) is challenging due to the variations in both the time- and frequency-domains. Against this background, the contribution of this paper is twofold: Firstly, we investigate the optimal sequence design criteria for efficient channel estimation in orthogonal frequency division multiplexing systems under DSC. Secondly, to design pilot sequences that can satisfy the derived criteria, we propose a new metric called oversampled ambiguity function (O-AF), which considers both fractional and integer Doppler frequency shifts. Optimizing the sidelobes of O-AF through a modified iterative twisted approximation (ITROX) algorithm, we develop a new class of pilot sequences called ``oversampled low ambiguity zone (O-LAZ) sequences". Through numerical experiments, we evaluate the efficiency of the proposed O-LAZ sequences over the traditional low ambiguity zone (LAZ) sequences, Zadoff-Chu (ZC) sequences and m-sequences, by comparing their channel estimation performances over DSC.
		\end{abstract}
		
		
		\section{Introduction}
There is a growing research interest in high-mobility and high-rate wireless communications, such as those utilized in high-speed train and Vehicle-to-Everything (V2X) communications, as well as low-earth-orbit satellite networks \cite{wu2016survey,ma2003optimal,leus,qu2008estimation,zilong2022}. These communications typically occur over doubly selective channels (DSC), which are challenging due to the time and frequency selectivity caused by multipath propagation and Doppler shifts/spread. Interference suppression in DSC is one of the main challenges faced by modern communication systems.

%There is a growing research interest in high-mobility and high-rate wireless communications, such as high-speed train and Vehicle-to-Everything (V2X) communications as well as low-earth-orbit satellite networks \cite{wu2016survey,ma2003optimal,leus,qu2008estimation,zilong2022}. Communications over these kinds of wireless channels, often called doubly selective channels (DSC), are challenging due to the time and frequency selectivity caused by multipath propagation and Doppler shifts/spread, respectively.
%Interference suppression in DSC is one of the main challenges faced by modern communication systems.

Orthogonal Frequency-Division Multiplexing (OFDM) systems have been successfully incorporated into important standards like the 3rd Generation Partnership Project (3GPP) Long Term Evolution (LTE), New Radio (NR), and IEEE 802.11a/g/n/ac/ax due to their excellent spectral efficiency, robustness to multipath fading, and low implementation complexity \cite{3gpp2019study,cho2010mimo}. However, in DSC, OFDM systems suffer from severe inter-carrier interference (ICI), which becomes more severe with increasing speed and carrier frequency. This interference negatively impacts the accuracy of channel estimation and the correctness of data demodulation. Extensive research has been conducted on channel estimation in DSC for OFDM systems.
%Orthogonal frequency-division multiplexing (OFDM) systems have been successfully included in important standards such as the 3rd Generation Partnership Project (3GPP) Long Term Evolution (LTE), New Radio (NR) and IEEE 802.11a/g/n/ac/ax due to their excellent spectral efficiency, robustness to multipath fading, and low implementation complexity \cite{3gpp2019study,cho2010mimo}.
%However, OFDM systems suffer from severe inter-carrier interference (ICI) in DSC, and the ICI becomes more severe with increasing speed and carrier frequency.
%In OFDM systems, ICI affects the accuracy of channel estimation and the correctness of data demodulation. There has been extensive research on channel estimation in DSC for OFDM systems.
For example, researchers have proposed conditions for the orthogonality between pilot and data symbols at the channel output \cite{kannu2005mse}. Based on this orthogonality condition, frequency-domain Kronecker delta (FDKD) pilots have been considered for frequency-domain pilot designs \cite{tugnait2002linear, tang2007pilot, islam2011optimum}. However, these schemes often require specific modifications to the OFDM frame structure, limiting their practical application. Although recursive channel estimation algorithms offer good ICI mitigation capabilities, they typically come with drawbacks such as high computational complexity and the need for detailed channel statistical information \cite{tugnait2009doubly, chopra2021data, liao2020nonlinear, mohebbi2021doubly}. These challenges make it difficult to apply these methods within LTE/NR standards.

In light of these challenges, achieving ICI suppression and accurate channel estimation with low complexity, without altering the existing LTE/NR frame structure, remains a hot research topic in the industry. This paper focuses on designing pilot sequences for channel estimation in DSCs to mitigate ICI, without requiring changes to the frame structure or channel estimation algorithms in LTE/NR. This approach enhances the robustness of the proposed sequences for practical engineering applications.

%\blue{The authors in \cite{kannu2005mse} proposed conditions for the orthogonality between pilot and data symbols at the channel output. Based on the orthogonality condition, the frequency-domain Kronecker delta (FDKD) family pilots are considered for frequency-domain pilot designs \cite{tugnait2002linear, tang2007pilot, islam2011optimum}.
%However, these schemes usually have special requirements regarding the OFDM frame structure.
%There are some recursive channel estimation algorithms. While they typically offer good ICI mitigation capabilities, they have drawbacks such as high computational complexity and the need for channel statistical information \cite{tugnait2009doubly, chopra2021data, liao2020nonlinear, mohebbi2021doubly}.
%In summary, these drawbacks make them difficult to apply in LTE/NR standards.
%Therefore, achieving ICI interference suppression and accurate channel estimation with low complexity within the existing LTE/NR frame structure is a hot research topic for the industry.

%In this paper, we focus on designing pilot sequences for channel estimation in DSCs to mitigate ICI, without changing the frame structure and channel estimation algorithms in LTE/NR.
%This makes the proposed sequences more robust for practical engineering applications.}

Zero/Low Correlation Zone (ZCZ/LCZ) sequences and Zero/Low Ambiguity Zone (ZAZ/LAZ) sequences are among the traditional sequences in sequence design. Sequences with good correlation properties have found many applications in modern communication systems \cite{chu,fan1996sequence,golomb2005signal}. In practice, sequences with zero correlation sidelobes in all delays are preferred, but achieving this is generally difficult. Given the quasi-synchronous nature of communication systems, the concept of ZCZ sequences was first introduced in 1999 \cite{fan1999class}, followed by LCZ sequences, where sidelobes within the LCZ region are bounded by a small positive value. Theoretical bounds for ZCZ and LCZ sequences under periodic correlation are derived in \cite{tang2000lower}.

%Zero/low correlation zone (ZCZ/LCZ) sequences, and zero/low ambiguity zone (ZAZ/LAZ) sequences are four traditional sequences in the field of sequence design.
%In the sequel, we will first review the state-of-the-art of ZCZ/LCZ sequences and ZAZ/LAZ sequences, then introduce our contributions.

%\subsubsection*{Zero/Low Correlation/Ambiguity Zone Sequence}
%Sequences with good correlation properties have found many applications in modern communication systems \cite{chu,fan1996sequence,golomb2005signal}. In practice, sequences with zero correlation sidelobes in all delays are preferred, but this is a difficult task in general.
%Considering the quasi-synchronous transmission nature of communication systems, the concept of ZCZ sequences was first proposed \cite{fan1999class} in 1999. Subsequently, LCZ sequences were proposed, whereby the sidelobes within the LCZ region are upper bounded by a small positive value. Theoretical bounds for ZCZ and LCZ sequences under periodic correlation are derived in \cite{tang2000lower}.

The relative movement between the transmitter and receiver causes signal distortion, known as the Doppler effect. The ambiguity function (AF) is used to measure Doppler frequency shifts and helps determine relative velocity in radar systems. An AF of a sequence $\mathbf{a}$, denoted by $AF_{\mathbf{a}}(\tau,f)$, is a two-dimensional function of the propagation delay $(\tau)$ and the Doppler shift ($f$) \cite{levanon}. The correlation of a sequence gives the AF of the sequence along the zero-Doppler axis. However, similar to the correlation function, it is impossible to maintain zero AF sidelobes for all non-zero Doppler shifts. Therefore, the concepts of ZAZ and LAZ sequences were introduced \cite{he2012waveform}, with theoretical bounds derived for these sequences under periodic correlation \cite{ye2022low}.

%It is well known that the relative movement between the transmitter and receiver causes signal distortion, a phenomenon known as the Doppler effect.
%\blue{The ambiguity function (AF) is used to measure Doppler frequency shifts caused by the Doppler effect, which in turn helps determine the relative velocity in radar.}
%An AF of a sequence $\mathbf{a}$, denoted by $AF_{\mathbf{a}}(\tau,f)$, is a two-dimensional function of the propagation delay $(\tau)$ and the Doppler shift ($f$) \cite{levanon}. The correlation of a sequence gives the AF of the sequence along the zero-Doppler axis. However, similar to the correlation function, it is impossible to maintain zero AF sidelobes for all non-zero Doppler shifts. Hence, similar to the concept of ZCZ/LCZ, zero ambiguity zone (ZAZ)/low ambiguity zone (LAZ) are introduced in \cite{he2012waveform}.
%Theoretical bounds for ZAZ and LAZ sequences under periodic correlation are derived in \cite{ye2022low}.

Systematic constructions of sequences with various correlation properties are typically based on algebraic tools \cite{fan1999class, deng2000spreading, Avik2020, zhi2023, tang2008new, tang2010multiple, Kaiqiang2024, Avik2022, ye2022low, gu2023asymptotically} or optimization algorithms. Algebraic constructions are straightforward to implement but often result in sequences with limited parameters. In contrast, optimization-based methods offer more flexibility \cite{he2012waveform, cui2020radar}, though they generally come with higher complexity. Notably, in 2004, Deng \cite{deng2004polyphase} proposed an algorithm that combined simulated annealing with heuristic search to suppress sidelobes of global aperiodic correlation, marking the start of sequence design based on algorithmic methods.

%Systematic constructions of sequences with various correlation properties are typically based on algebraic tools \cite{fan1999class, deng2000spreading, torii2004new, zhi2023, tang2008new, tang2010multiple, liu2013new, Avik2022, ye2022low, gu2023asymptotically}, or using optimization algorithms. Algebraic constructions are easy to implement, but they often lead to sequences with limited parameters. Constructions based on optimization algorithms, on the other hand, are relatively flexible \cite{he2012waveform, cui2020radar}.
%In 2004, Deng proposed an algorithm to suppress sidelobes of global aperiodic correlation based on a  hybrid approach combining simulated annealing algorithm \cite{deng2004polyphase}.
%Although this work utilizes a heuristic search algorithm rather than an optimization algorithm, it ushered in the era of sequence design based on algorithmic methods.

Due to the high complexity of heuristic search algorithms, researchers have developed new sequence generation algorithms based on optimization theory. Examples include the Cyclic-Algorithm-Original (CAO), proposed in 2008, aiming to reduce periodic/aperiodic correlation sidelobes through SVD decomposition of the correlation matrix \cite{li2008signal}. Subsequently Cyclic-algorithm-new (CAN) algorithm \cite{stoica2009new, stoica2009designing}, based on alternating minimization techniques, was designed to synthesize unimodular aperiodic sequences of large length. However, its drawback is that the solution of the approximate problem does not always converge to the minimum of the original problem.
The Monotonic minimizer for integrated sidelobe level (MISL) \cite{song2015optimization} algorithm utilizes the majorization-minimization (MM) method to minimize a surrogate function of the original integrated sidelobe level (ISL) minimization problem, ensuring convergence to the global minima of the original problem. Nevertheless, its convergence is slow due to the double majorization of the original cost function. In recent years, efficient algorithms for generating unimodular sequences have emerged, such as Limited-Memory BFGS (LBFGS) \cite{nocedal1999numerical, wang2012design}, Iterative Twisted Approximation (ITROX) computational framework \cite{soltanalian2012computational, gu2022computational, gu2023computational}, Power Spectral Density Fitting-based Iterative Approach (PIA) \cite{wu2015designing}, Efficient Gradient Descent (EGD) \cite{baden2017multiobjective}, Coordinate Descent (CD) \cite{kerahroodi2017coordinate}, Alternating Direction Method of Multipliers (ADMM), Parallel Direction Method of Multipliers (PDMM) \cite{wang2021designing}, algorithm based on neural networks \cite{rezaei2023learning}. Most of the algorithms mentioned above focus on achieving better suppression levels for correlation sidelobes than the earlier methods could achieve \cite{liang2015waveform, zhao2016unified, esmaeili2016designing, li2017fast, lin2019efficient}.
Due to space limitation, these methods are not discussed in details here.
		In Table \ref{table_corr}, we give an overview of relevant sequence design algorithms that can design sequences with low correlations, for comparison.
		Here, we only present the complexity of the correlation optimization problem for a single sequence of length $N$ and LCZ width $Z$.
		\begin{table}[ht]
			\centering
			\caption{Comparison of LCZ Sequence Design Algorithms}\label{table_corr}
			\begin{tabular}{|c|c|c|c|}
				\hline
				Algorithms & \makecell{Computational\\Complexity} & \makecell{Aperiodic\\ Correlation} & \makecell{Periodic\\ Correlation} \\ \hline
				CAO \cite{li2008signal} & $\mathcal{O}(NZ^2+Z^3)$ & $\surd$ & $\surd$ \\
				CAN \cite{stoica2009new} & $\mathcal{O}(N^2)$ & $\surd$ & $\times$ \\
				PeCAN \cite{stoica2009designing} & $\mathcal{O}(N^2)$ & $\times$ & $\surd$ \\
				LBFGS \cite{wang2012design} & $\mathcal{O}(NZ+N\log_2 N)$ & $\surd$ & $\times$ \\
				ITROX \cite{soltanalian2012computational} & $\mathcal{O}(N^3)$ & $\surd$ & $\surd$ \\
				PIA \cite{wu2015designing} & $\mathcal{O}(N^2)$ & $\surd$ & $\times$ \\
				MISL \cite{song2015optimization} & $\mathcal{O}(N^2)$ & $\surd$ & $\surd$ \\
				EGD \cite{baden2017multiobjective} & $\mathcal{O}(N^2)$ & $\surd$ & $\times$ \\
				\hline
			\end{tabular}
		\end{table}

Similar to the generation of ZCZ/LCZ sequences, optimization-based algorithms are the primary methods for generating ZAZ/LAZ sequences. Additionally, some ZCZ/LCZ sequence generation methods can be adapted to improve ZAZ/LAZ sequence generation algorithms, such as the AF-CAO \cite{li2008signal}.
Besides, there are many other optimization methods such as Maximum Block Improvement (MBI) method \cite{aubry2013ambiguity}, gradient descent (GD) \cite{arlery2016efficient}, coordinate iteration for ambiguity function iterative shaping (CIAFIS) \cite{wu2016cognitive}, accelerated iterative sequential optimization (AISO) algorithm \cite{cui2017local, fu2017ambiguity}, Lagrange programming neural network-alternating direction method of multipliers (LPNN-ADMM) \cite{jing2018designing}, quartic GD \cite{alhujaili2019quartic}, projected GD, manifold optimization embedding with momentum (MOEM), LBFGS, etc \cite{esmaeili2019unimodular, wei2023unimodular, hu2023ambiguity}. Although GD-based algorithms are generally effective, they tend to be slow due to the need to calculate gradients at every step.
In Table \ref{table_AF}, we give an overview of relevant sequence design algorithms that can design sequences with low ambiguity, for comparison.
Again, we only present the complexity of the ambiguity optimization problem for a single sequence of length $N$ and LAZ size $Z\times F$.
		\begin{table}[ht]
			\centering
			\caption{Comparison of LAZ Sequence Design Algorithms}\label{table_AF}
			{
				\begin{tabular}{|c|c|c|c|}
					\hline
					Algorithms & \makecell{Computational\\Complexity} & \makecell{Aperiodic\\AF} & \makecell{Periodic\\AF} \\ \hline
					AF-CAO \cite{li2008signal} & $\mathcal{O}((N+Z)Z^2F^2)$ & $\surd$ & $\surd$ \\
					MBIL \cite{aubry2013ambiguity} & $\mathcal{O}(N)$ & $\surd$ & $\times$ \\
					MBIQ \cite{aubry2013ambiguity} & $\mathcal{O}(N^{3.5})$ & $\surd$ & $\times$ \\
					GD \cite{arlery2016efficient} & $\mathcal{O}(N^{2}F)$ & $\surd$ & $\surd$ \\
					CIAFIS \cite{wu2016cognitive} & $\mathcal{O}(N^{2})$ & $\surd$ & $\times$ \\
					AISO \cite{cui2017local, fu2017ambiguity} & Not report & $\surd$ & $\times$ \\
					LPNN-ADMM \cite{jing2018designing} & $\mathcal{O}(N^{2}ZF)$ & $\surd$ & $\times$ \\
					\hline
			\end{tabular}}
		\end{table}

%Overall, the above results on ZCZ/LCZ/ZAZ/LAZ sequences offer the advantages of flexible length and flexible ZCZ and ZAZ sizes.

These ZCZ/LCZ/ZAZ/LAZ sequences offer the advantages of flexible length and adaptable ZCZ and ZAZ sizes, making them suitable for various applications, including radar waveform design \cite{li2008mimo, li2008signal}. Channel estimation, which involves separating time delay and Doppler shift, is conceptually similar to radar's velocity and range measurement. Motivated by this similarity, this paper explores the use of ZAZ/LAZ sequences as pilot sequences for channel estimation in DSC.

%\blue{
%The AF plays a crucial role in radar waveform design. ZAZ and LAZ sequences have excellent velocity and range measurement capabilities \cite{li2008mimo, li2008signal}. Channel estimation is quite similar to velocity and range measurement in radar, as both tasks involve separating the time delay and Doppler shift of each path. Motivated by this idea, we aim to use ZAZ/LAZ sequences as pilot sequences for channel estimation in DSC.
%}

\subsubsection*{Contributions}

To develop efficient preamble sequences in OFDM, we first analyze the requirements for pilot sequences in DSC channel estimation. Based on the Jakes’ model of the Rayleigh fading channel (to be detailed in Section \ref{sec4A}), we observe that Doppler shifts in DSC can be non-integer, which limits the effectiveness of conventional ZAZ and LAZ sequences. To address this issue, we introduce the concept of oversampled ambiguity functions (O-AF) for more accurate estimation of channel responses at both integer and fractional Doppler shifts. This concept leads to the development of oversampled zero/low ambiguity zone (O-ZAZ/O-LAZ) sequences.

%Aiming to develop efficient preamble sequences in OFDM, we first analyze the channel estimation requirements of pilot sequences in DSC. Based on the Jakes' model (to be detailed in Section \ref{sec4A}) of the Rayleigh fading channel, we observe that the Doppler shifts in DSC can be non-integer. In this case, conventional ZAZ and LAZ sequences are not effective as their zero/low ambiguity sidelobes may not be guaranteed over fractional delay-Doppler scenarios.
		
%To address this problem, we introduce the concept of oversampled AF (O-AF) for accurate estimation of the channel responses at integer delay and fractional Doppler shifts. Traditional AF is similar to O-AF, but the only difference is that the Doppler axis in O-AF is oversampled and this helps us to obtain the magnitude of the AF at both integer and fractional Doppler shifts. Such a feature is very useful in matching the non-integer Doppler shifts in DSC, thus leading to improved channel estimation.

%Through trial and error, we modify the ITROX algorithm to design O-LAZ sequences, naming this modified version the oversampled ambiguity ITROX (OA-ITROX) algorithm. The OA-ITROX algorithm is capable of designing both traditional LAZ and O-LAZ sequences. We further address the complexity of OA-ITROX by employing the power method for rank-1 approximation instead of SVD. Finally, numerical experiments demonstrate that sequences with low O-AF properties can be effectively used as pilot sequences for channel estimation in DSC.

Similar to traditional ZAZ/LAZ, we coin the concept of oversampled zero/low ambiguity zone (O-ZAZ/O-LAZ) if there is a region in the ambiguity plot with zero/low sidelobes. Through a series of trial and error (as detailed in \textit{Remark} \ref{r6}), we chose the traditional ITROX algorithm and modify it to design O-LAZ sequences. We named the modified ITROX algorithm as oversampled ambiguity ITROX (OA-ITROX) algorithm. The major difference between OA-ITROX with the traditional ITROX is that, here we incorporate submatrices with frequency offsets and have replaced eigenvalue decomposition (EVD) with SVD. The proposed OA-ITROX algorithm is capable of designing both traditional LAZ as well as O-LAZ sequences. It is shown that the complexity of OA-ITROX highly depends on SVD. We further address the complexity of OA-ITROX by employing the power method for rank-1 approximation instead of SVD (as shown in \textit{Remark} \ref{rem5}). Finally, numerical experiments demonstrate that sequences with low O-AF properties can be effectively used as pilot sequences for channel estimation in DSC.

\subsubsection*{Organization}
The rest of the paper is organized as follows. Section II introduces the notations and discusses ZCZ/LCZ and ZAZ/LAZ sequences, followed by an overview of OFDM system principles and channel estimation. In Section III, we revisit the Jakes’ model and analyze its impact on signals, leading to the derivation of design criteria for pilot sequences in DSC. Section IV presents the OA-ITROX algorithm and the designed O-LAZ sequences. Section V evaluates the proposed training scheme through simulations. Finally, Section VI concludes the paper.

		\section{Preliminaries}\label{sec2}
%		\subsection{Notations}
		The following notations will be used throughout this paper.
		\begin{itemize}
			\item $\mathbf{X}^*, \mathbf{X}^\mathrm{T}$ and $\mathbf{X}^\mathrm{H}$ denote the complex conjugate, the transpose and the conjugate transpose of matrix $\mathbf{X}$, respectively;
			\item $\langle\mathbf{a},\mathbf{b}\rangle$ denotes the inner-product between two complex valued sequences $\mathbf{a}=[a[0],a[1],\ldots,a[N-1]]^\mathrm{T}$, \\ $\mathbf{b}=[b[0],b[1],\ldots,b[N-1]]^\mathrm{T}$, i.e., $\langle\mathbf{a},\mathbf{b}\rangle= \sum_{k=0}^{N-1}a[k]b^*[k]$, where $N$ is the sequence length of $\mathbf{a}$ (and $\mathbf{b}$);
			\item $[\mathbf{a}]_{k}$ denotes the $k$-th element of sequence $\mathbf{a}$;
			\item $[\mathbf{X}]_{i,j}$ denotes the $i$-th row $j$-th column element of matrix $\mathbf{X}$;
			\item $S^\tau(\mathbf{a})$ denotes the right-cyclic-shift of $\mathbf{a}=[a[0],a[1],\ldots,a[N-1]]^\mathrm{T}$ for $\tau$ (nonnegative integer) positions, i.e.,
$$
				S^\tau(\mathbf{a})= 
				[\underbrace{a[N-\tau], \ldots, a[N-1]}_{\text {the last } \tau \text { elements of } \mathbf{a}}, a[0], a[1], \ldots, a[N-\tau-1]]^{\mathrm{T}};
$$
			Similarly,
$$
				S^{-\tau}(\mathbf{a})= 
				[a[\tau], a[\tau+1], \ldots, a[N-1], \underbrace{a[0], a[1], \ldots, a[\tau-1]}_{\text {the first } \tau \text { elements of } \mathbf{a}}]^{\mathrm{T}};
$$
			\item $\lfloor n\rfloor_N$ denotes $n \pmod N$;
			\item $\zeta_N$ denotes the $N$-th complex roots of unity, {i.e., $\zeta_N=e^{2\pi i/N}$;}
			\item $\mathbf{F}_N$ denotes the Fourier matrix of size $N$, i.e., $[\mathbf{F}_N]_{i,j}=\frac{1}{\sqrt{N}}\zeta_N^{-(i-1)(j-1)}$;
			\item $\mathbf{a}\odot\mathbf{b}$ denotes elementwise multiplication, i.e., $[\mathbf{a}\odot\mathbf{b}]_k=a[k]b[k]$;
			\item $\mathbf{a}\oslash\mathbf{b}$ denotes elementwise division, i.e., $[\mathbf{a}\oslash\mathbf{b}]_k=a[k]/b[k]$;
			\item $\mathrm{mean}(\mathbf{a})$ denotes the mean of $\mathbf{a}$, i.e., $\mathrm{mean}(\mathbf{a})=(\sum_{k=0}^{N-1} a[k])/N$;
			\item $\varphi(\mathbf{a})$ denotes the phase of each of the elements of $\mathbf{a}$;
		\item $E(\cdot)$ denotes the expected value of a random variable.
		\end{itemize}
		
		For two length-$N$ complex-valued sequences $\mathbf{a}=[a[0],a[1],\ldots,a[N-1]]^\mathrm{T}$, $\mathbf{b}=[b[0],b[1],\ldots,b[N-1]]^\mathrm{T}$, $\phi_{\mathbf{a},\mathbf{b}}(\tau)$ denotes the periodic cross-correlation function (PCCF) between $\mathbf{a}$ and $\mathbf{b}$, i.e.,
		\begin{equation}\label{equation_PCCF}
			\phi_{\mathbf{a},\mathbf{b}}(\tau)= \sum_{k=0}^{N-1} a[k]b^*[\lfloor k+\tau\rfloor_N]=\langle\mathbf{a},S^{-\tau}(\mathbf{b})\rangle.
		\end{equation}
		In particular, when $\mathbf{a}=\mathbf{b}$, $\phi_{\mathbf{a},\mathbf{b}}(\tau)$ is written as $\phi_{\mathbf{a}}(\tau)$ and called the periodic auto-correlation function (PACF) of $\mathbf{a}$ at time-shift $\tau$.
		
		For two length $N$ complex-valued sequences $\mathbf{a}=[a[0],a[1],\ldots,a[N-1]]^\mathrm{T}$, $\mathbf{b}=[b[0],b[1],\ldots,b[N-1]]^\mathrm{T}$, $AF_{\mathbf{a},\mathbf{b}}(\tau,f)$ denotes the periodic cross-ambiguity function (PCAF) between $\mathbf{a}$ and $\mathbf{b}$ at time-shift $\tau$ and frequency-shift $f$, i.e.,
		\begin{equation}\label{equation_PCAF}
			AF_{\mathbf{a},\mathbf{b}}(\tau,f)= \sum_{k=0}^{N-1} a[k]b^*[\lfloor k+\tau\rfloor_N]\zeta_N^{fk} =\langle\mathbf{a}\odot\mathbf{f},S^{-\tau}(\mathbf{b})\rangle,
		\end{equation}
		where $\mathbf{f}=[1,\zeta_N^{f},\ldots,\zeta_N^{f(N-1)}]^\mathrm{T}$.
		In particular, when $\mathbf{a}=\mathbf{b}$, $AF_{\mathbf{a},\mathbf{b}}(\tau,f)$ is written as $AF_{\mathbf{a}}(\tau,f)$ and called the periodic auto-ambiguity function (PAAF) of $\mathbf{a}$ at time-shift $\tau$ and frequency-shift $f$.
		
		\begin{Remark}
			Note that in (\ref{equation_PCCF}) and (\ref{equation_PCAF}), the values of $\tau$ and $f$ are integers, i.e., $\tau, f = 0, 1, \ldots, N-1$.
		\end{Remark}
		
		\subsection{Zero/Low Correlation/Ambiguity Zone Sequence}
		Next, we give a brief introduction to zero/low correlation/ambiguity zone sequence.
		Integrated sidelobe level (ISL) is an important metric, which is used to characterize the correlation and AF of sequences.
		\begin{Definition}
			Let $\mathbf{a}=[a[0],a[1],\ldots,a[N-1]]^\mathrm{T}$ be a sequence of length $N$, the ISL of correlation is define by
			\begin{equation}\label{}
				\mathrm{ISL}_\phi(\mathbf{a};Z)=2\sum_{\tau=1}^{Z} |\phi_{\mathbf{a}}(\tau)|^2,
			\end{equation}
			where $Z$ is the size of concerned correlation zone.
			Similarly, the ISL of AF is define by
			\begin{equation}\label{}
				\mathrm{ISL}_{AF}(\mathbf{a};Z\times F_r)=2\sum_{\tau=1}^{Z}\sum_{f=-F_r}^{F_r} |AF_{\mathbf{a}}(\tau,f)|^2,
			\end{equation}
			where $Z\times F_r$ is the size of concerned ambiguity zone.
		\end{Definition}
		
		By the above definition of ISL, we can give the definitions of zero correlation/ambiguity zone sequence.
		\begin{Definition}
			A sequence $\mathbf{a}$ of length $N$ is called a zero correlation zone (ZCZ) sequence with ZCZ width $Z$, if
			\begin{equation}\label{equation_ZCZ}
				\mathrm{ISL}_\phi(\mathbf{a};Z)=0.
			\end{equation}
			Similarly, $\mathbf{a}$ is called a zero ambiguity zone (ZAZ) sequence with ZAZ size $Z\times F_r$, if
			\begin{equation}\label{equation_ZAZ}
				\mathrm{ISL}_{AF}(\mathbf{a};Z\times F_r)=0.
			\end{equation}
		\end{Definition}
		Designing sequences which satisfy conditions (\ref{equation_ZCZ}) and (\ref{equation_ZAZ}) are very challenging.
		Low correlation/ambiguity zone is generalized version of ZCZ/ZAZ. Normally, it is required that ISL should be as low as possible.
		\begin{Definition}
			A sequence $\mathbf{a}$ of length $N$ is called a low correlation zone (LCZ) sequence with LCZ width $Z$, if
			\begin{equation}\label{equation_LCZ}
				\mathrm{ISL}_\phi(\mathbf{a};Z)<\varepsilon,
			\end{equation}
			where $\varepsilon$ is a very small value.
			
			Similarly, $\mathbf{a}$ is called a low ambiguity zone (LAZ) sequence with LAZ size $Z\times F_r$, if
			\begin{equation}\label{equation_LAZ}
				\mathrm{ISL}_{AF}(\mathbf{a};Z\times F_r)<\varepsilon.
			\end{equation}
		\end{Definition}
		
		\subsection{System Model of OFDM and Channel Estimation}\label{sec2b}
\begin{figure}[ht]
			\centering
			\includegraphics[width=8cm]{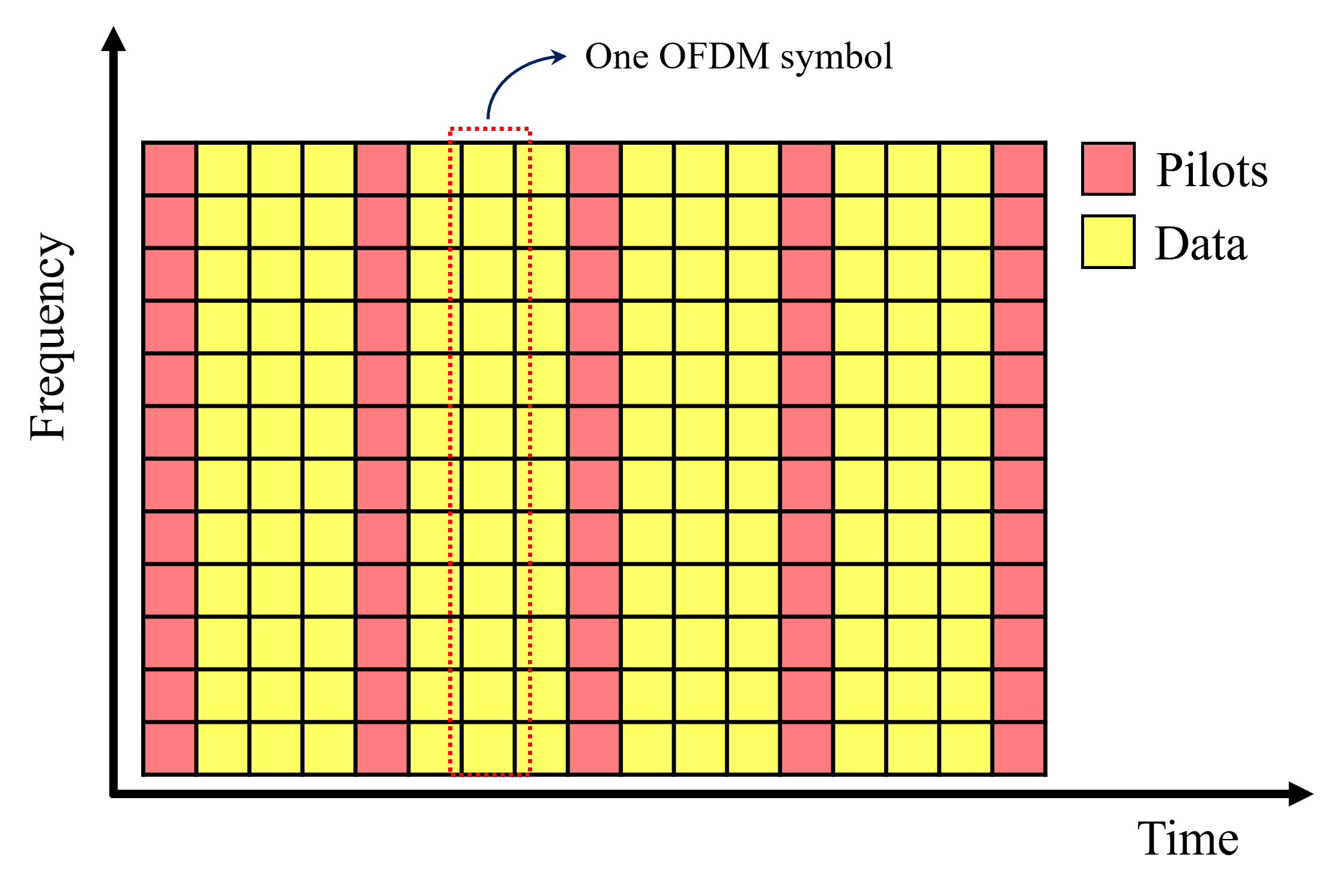}
			\caption{The transmission frame structure.}\label{fig_frame}
		\end{figure}

		In this paper, we consider an OFDM transmission system with $N$ carriers. To simplify the system model, we assume that both the transmitter and receiver have a single antenna.
		Fig. \ref{fig_frame} shows the transmission frame structure of the OFDM system.
		Within a frame, we assume a total of $M_P$ pilot OFDM symbols, with each pair of pilot OFDM symbols containing $M_D$ data OFDM symbols. Consequently, an OFDM frame comprises a total of $M=M_P+M_D(M_P-1)$ OFDM symbols.
		
		For the $m$-th OFDM symbol, the time-domain received signal after removing the cyclic prefix can be represented as $\mathbf{y}_m=[y_{m}[0],y_{m}[1],\ldots,y_{m}[N-1]]^\mathrm{T}, m=0,1,\ldots,M-1$, and it can be expressed as:
		$$y_{m}[n]=\sum_{l=0}^{L-1}h_m[l,n]x_m[\lfloor n-l\rfloor_N]+w_m[n],$$
		where $\mathbf{x}_m=[x_m[0],x_m[1],\ldots,x_m[N-1]]^\mathrm{T}$ represents the $m$-th OFDM symbol in the time-domain. $h_m[l,n]$ is the channel response matrix of $m$-th OFDM symbol in the time-domain.
		%The frequency-domain received model for the $m$-th OFDM symbol can be represented as:
		%$$\mathbf{Y}_m=\mathbf{H}_m\mathbf{X}_m+\mathbf{W}_m,$$
		%where $\mathbf{X}_m$is the transmitted signal matrix in the frequency domain for the $m$-th OFDM symbol, $\mathbf{H}_m =\mathbf{F}_N\mathbf{H}_{\mathrm{T}m}\mathbf{F}_N^H, [\mathbf{H}_{\mathrm{T}m}]_{p,q}=h_m[p,\lfloor p-q\rfloor_N]$.
		Channel estimation is first performed over pilot OFDM symbols and then discrete prolate spheroidal sequences (DPSS) interpolation is used to achieve channel estimation over the data symbols \cite{zemen2005time}.
		%	\textcolor{red}{It is common to first} perform channel estimation for a fixed time of the pilot OFDM symbols and then use discrete prolate spheroidal sequences (DPSS) interpolation to achieve channel estimation for the data symbols \cite{zemen2003time, zemen2005time}.
		%	Therefore, accurately performing channel estimation for the pilot OFDM symbols becomes a primary task.
		
		%	The LS assessment strategy is a valuable method for channel estimation techniques.
		Without loss of generality, we consider the channel estimation model for a single pilot OFDM symbol by least square (LS) method.
		Let $\mathbf{x}=\mathbf{a}$ be the time-domain sequence transmitted as the pilot OFDM symbols, then the LS estimation is given by
		\begin{equation}\label{equation_LS}
			\hat{\mathbf{h}}=(\mathbf{A}_T^H\mathbf{A}_T)^{-1}\mathbf{A}_T^H\mathbf{y},
		\end{equation}
		where $\mathbf{y}$ is the time-domain sequence received after removing the cyclic prefix, $\mathbf{A}_T$ is the cyclic matrix of size $N\times L$ ($L$ is the maximum multipath delay) generated by $\mathbf{a}$, and it is defined as follows
		
			\begin{equation}\label{equation_AT}
				\mathbf{A}_T=
				\left[
				\begin{array}{cccc}
					a[0] & a[N-1] & \cdots & a[N-L+1] \\
					a[1] & a[0] & \cdots & a[N-L+2] \\
					\vdots & \vdots &   & \vdots \\
					a[N-1] & a[N-2] & \cdots & a[N-L] \\
				\end{array}
				\right].
			\end{equation}

		From (\ref{equation_LS}), it can be observed that the channel estimation obtained by the LS method is a vector, representing an estimate of the true channel response at a certain instant. Mostofi \textit{et al.} \cite{mostofi2005ici} proved that the channel estimated by the LS method corresponds to the estimate of the midpoint instant of the true channel response in the sense of statistical averaging \cite{mostofi2005ici}.
		\begin{Lemma}[\cite{mostofi2005ici}]
			Let $h[l,n]$ be the true channel response at the $n$-th instant and the $l$-th path corresponding to an OFDM symbol, and $\mathbf{h}_n$ contains the channel response of all the $L$ multipaths at $n$-th instant, i.e., $\mathbf{h}_n=[h[0,n],h[1,n],\ldots,h[L-1,n]]^\mathrm{T}$.
			The average channel response over the OFDM time duration of $n = [0,1,\ldots,N-1]$ is closest to the midpoint instant $\mathbf{h}_\mathrm{mid}=\mathbf{h}_{\frac{N-1}{2}}$.
			This can be mathematically described as follows
			$$
			E\left(\|\mathbf{h}_\mathrm{avg}- \mathbf{h}_n \|^2\right) \hbox{ is minimized for } n= \frac{N-1}{2},
			$$
			where $\mathbf{h}_\mathrm{avg}=1/N\sum_{n=0}^{N-1}\mathbf{h}_n$.
			Furthermore, the channel estimation values $\hat{\mathbf{h}}$ obtained by the LS method are estimated of the average channel response, i.e.,
			$\hat{\mathbf{h}}=\hat{\mathbf{h}}_\mathrm{avg}$.
			And it is also close to the midpoint instant of actual channel response,
			$$
			E\left(\|\hat{\mathbf{h}}- \mathbf{h}_n \|^2\right) \hbox{ is minimized for } n= \frac{N-1}{2}.
			$$
		\end{Lemma}

\section{Optimal Pilot Design Criterion for DSC}
Jakes model has become a well-established benchmark in the research community due to its realistic Doppler spectrum representation, which accurately models the frequency variations caused by the relative motion between the transmitter and receiver \cite{zheng2003simulation}.
In this section, we first revisit the Jakes' channel model for DSC. Then we derive the optimal pilot design criterion from the channel model.

		\subsubsection*{Jakes' Model for DSC}\label{sec4A}
		Here we consider the case of a single OFDM symbol with $N$ subcarriers. The time-domain received sequence after removing the cyclic prefix can be represented as $\mathbf{y}=[y[0],y[1],\ldots,y[N-1]]^\mathrm{T}$ which can be expressed as follows:
		\begin{equation}\label{equation_channel}
			y[n]=\sum_{l=0}^{L-1}h[l,n]x[\lfloor n-l\rfloor_N]+w[n],
		\end{equation}
		where $L$ denotes the number of delay paths, $\mathbf{x}=[x[0],x[1],\ldots,x[N-1]]^\mathrm{T}$ represents the time-domain transmitted sequence, and $h[l,n]$ is an element of time-domain channel response matrix, which represents the channel at the $n$-th sampling time on the $l$-th multipath.
The improved Jakes' simulator \cite{zheng2003simulation} by sum-of-sinusoids statistical simulation models is given by
		\begin{equation}\label{equation_Jakes}
			h[l,t]=E_0\sum_{q=0}^{Q-1} A_l C_q^{(l)} e^{i(2\pi f_d t \cos \alpha_q^{(l)}+\phi_q^{(l)})},
		\end{equation}
		where
		\begin{align*}
			& C_q^{(l)}=\frac{e^{i\psi_q^{(l)}}}{\sqrt{Q}}, q=0,1,\ldots,Q-1, \\
			& \alpha_q^{(l)}=\frac{2\pi q-\pi+\theta^{(l)}}{Q}, q=0,1,\ldots,Q-1, \\
			& \phi_q^{(l)}=\phi_{Q/2+q}^{(l)}=\tilde{\phi}_q^{(l)}, q=0,1,\ldots,Q/2-1, \\
		\end{align*}
		$Q$ denotes the number of Doppler paths of each delay path and it is assumed to be even, $E_0$ the scaling constant, $f_d$ the maximum Doppler frequency, $A_l$ denotes the power of the the $l$-th path and follows complex Gaussian distribution with a mean of zero and a variance of $\sqrt{P_l}$, and $\psi_q^{(l)}, \theta^{(l)}$ and $\tilde{\phi}_q^{(l)}$ being mutually independent random variables uniformly distributed over $[-\pi,\pi)$.
		
		Note that in OFDM, the received signal is discretely sampled from the interval $[0, T_s]$, i.e., $t = nT_s/N$, $n=0,1,\ldots,N-1$, where $T_s$ denotes the OFDM symbol duration.
		Let $F_r$ denote the normalized maximum Doppler frequency, i.e., $F_r=f_d/\Delta f=f_d T_s$, then (\ref{equation_Jakes}) can be written as
		\begin{align}\label{equation_Jakes2}
			h[l,n]&=E_0\sum_{q=0}^{Q-1} A_l C_q^{(l)} e^{i(2\pi F_r n/N \cos \alpha_q^{(l)}+\phi_q^{(l)})} \notag \\
			&= E_0\sum_{q=0}^{Q-1} A_l C_q^{(l)} e^{i\phi_q^{(l)}} e^{2\pi i /N(F_r n \cos \alpha_q^{(l)})} \notag \\
			&= E_0\sum_{q=0}^{Q-1} A_l C_q^{(l)} e^{i\phi_q^{(l)}} \zeta_N^{F_r \cos \alpha_q^{(l)} n}.
		\end{align}
		
		By aggregating the random variables and neglecting the impact of large-scale fading, (\ref{equation_Jakes2}) becomes
		\begin{equation}\label{equation_Jakes3}
			h[l,n]= \sum_{q=0}^{Q-1} g_q^{(l)}\zeta_N^{\omega_q^{(l)}n},
		\end{equation}
		where $g_q^{(l)}=A_l C_q^{(l)} e^{i\phi_q^{(l)}}$, and $\omega_q^{(l)}=F_r \cos \alpha_q^{(l)}$. Note that from (\ref{equation_Jakes3}), it can be observed that the range of $\omega_q^{(l)}$ is in $[-F_r, F_r]$, and they are independently and identically distributed.
		
		\subsection{Optimal Pilot Design Criterion}\label{sec4a}
		In this subsection, we derive the optimal pilot design criterion over DSC.
		Substituting (\ref{equation_Jakes3}) in (\ref{equation_channel}), we obtain the transmission model for OFDM as follows
		\begin{equation}\label{equation_channel2}
			y[n]=\sum_{l=0}^{L-1} \sum_{q=0}^{Q-1} g_q^{(l)} \zeta_N^{\omega_q^{(l)} n} x[\lfloor n-l\rfloor_N]+w[n].
		\end{equation}
		
		{
			Let $\mathbf{x}=\mathbf{a}$ be the pilot sequence. For the ease of analysis, (\ref{equation_channel2}) can be represented in matrix form as follows:
			\begin{equation}\label{equation_channel3}
				\mathbf{y}=\sum_{q=0}^{Q-1}  (\mathbf{A}_T \odot \mathbf{\Omega}_q) \mathbf{g}_q+\mathbf{w},
			\end{equation}
			where $\mathbf{g}_q=[g_q^{(0)},g_q^{(1)},\ldots,g_q^{(L-1)}]^\mathrm{T}$, $\mathbf{A}_T$ is the cyclic matrix of size $N\times L$ generated by $\mathbf{a}$, as given in (\ref{equation_AT}) and
			$\mathbf{\Omega}_q$ is the Vandermonde matrix of size $N\times L$ generated by $\omega_q^{(l)}$, as follows:
			$$
			\mathbf{\Omega}_q=
			\left[
			\begin{array}{cccc}
				\zeta_N^{\omega_q^{(0)}0} & \zeta_N^{\omega_q^{(1)}0} & \cdots & \zeta_N^{\omega_q^{(L-1)}0} \\
				\zeta_N^{\omega_q^{(0)}1} & \zeta_N^{\omega_q^{(1)}1} & \cdots & \zeta_N^{\omega_q^{(L-1)}1} \\
				\vdots & \vdots &   & \vdots \\
				\zeta_N^{\omega_q^{(0)}(N-1)} & \zeta_N^{\omega_q^{(1)}(N-1)} & \cdots & \zeta_N^{\omega_q^{(L-1)}(N-1)} \\
			\end{array}
			\right].
			$$
Then the LS estimation is given by}
		\begin{align}\label{equation_LS2}
			\hat{\mathbf{h}} &=(\mathbf{A}_T^\mathrm{H}\mathbf{A}_T)^{-1}\mathbf{A}_T^\mathrm{H}\mathbf{y} \notag\\
			&= (\mathbf{A}_T^\mathrm{H}\mathbf{A}_T)^{-1}\mathbf{A}_T^\mathrm{H} \left(\sum_{q=0}^{Q-1}  (\mathbf{A}_T \odot \mathbf{\Omega}_q) \mathbf{g}_q +\mathbf{w} \right) \notag \\
			&= \sum_{q=0}^{Q-1} (\mathbf{A}_T^\mathrm{H}\mathbf{A}_T)^{-1} \mathbf{A}_T^\mathrm{H}(\mathbf{A}_T \odot \mathbf{\Omega}_q) \mathbf{g}_q +(\mathbf{A}_T^\mathrm{H}\mathbf{A}_T)^{-1} \mathbf{A}_T^\mathrm{H}\mathbf{w}.
		\end{align}
		
		Using \textit{Lemma 1}, $\hat{\mathbf{h}}$ is an estimate of $\mathbf{h}_\mathrm{avg}$. The corresponding MSE is
		\begin{equation}\label{equation_MSE}
			\mathrm{MSE}_\mathrm{avg}=E(\|\hat{\mathbf{h}}-\mathbf{h}_\mathrm{avg}\|^2).
		\end{equation}
		
		By substituting (\ref{equation_LS2}) into (\ref{equation_MSE}), we can derive the intrinsic relationship between MSE and the pilot sequence, given in  (\ref{equation_MSE2}),

			\begin{align}\label{equation_MSE2}
				\mathrm{MSE}_\mathrm{avg} &= E\left(\left\|\sum_{q=0}^{Q-1} (\mathbf{A}_T^\mathrm{H}\mathbf{A}_T)^{-1} \mathbf{A}_T^\mathrm{H}(\mathbf{A}_T \odot \mathbf{\Omega}_q) \mathbf{g}_q +(\mathbf{A}_T^\mathrm{H}\mathbf{A}_T)^{-1} \mathbf{A}_T^\mathrm{H}\mathbf{w}-\mathbf{h}_\mathrm{avg}\right\|^2\right) \notag\\
				&= E\left(\left\|\sum_{q=0}^{Q-1} \left[(\mathbf{A}_T^\mathrm{H}\mathbf{A}_T)^{-1} \mathbf{A}_T^\mathrm{H}(\mathbf{A}_T \odot \mathbf{\Omega}_q) -\mathrm{diag}(\mathbf{\Gamma}_q)\right]\mathbf{g}_q +(\mathbf{A}_T^\mathrm{H}\mathbf{A}_T)^{-1} \mathbf{A}_T^\mathrm{H}\mathbf{w}\right\|^2\right).
			\end{align}
		where
		\begin{equation}\label{equation_Gamma}
			\mathbf{\Gamma}_q=
			\left[
			\begin{array}{c}
				\frac{1}{N}\sum_{n=0}^{N-1} \zeta_N^{w_q^{(0)}n} \\
				\frac{1}{N}\sum_{n=0}^{N-1} \zeta_N^{w_q^{(1)}n} \\
				\vdots \\
				\frac{1}{N}\sum_{n=0}^{N-1} \zeta_N^{w_q^{(L-1)}n} \\
			\end{array}
			\right].
		\end{equation}
		
		From traditional channel estimation theory, we know that the matrix $\mathbf{A}_T^\mathrm{H}\mathbf{A}_T$ is required to be an identity matrix of order $N$ \cite{kay1993fundamentals} for minimum MSE.
		Hence, we get the first criterion ($\mathrm{CR}_1$) of pilot design over DSC as follows:
		\begin{equation}\label{Opt_criterion1}
			\arg\min_\mathbf{a}\|\mathbf{A}_T^\mathrm{H}\mathbf{A}_T-N\mathbf{I}_L\|_F^2,
		\end{equation}
		where $\mathbf{I}_L$ is an identity matrix of size $L\times L$.
		
		Furthermore, from (\ref{equation_MSE2}), the LS estimator is unbiased if the pilot sequence satisfies
		$$(\mathbf{A}_T^\mathrm{H}\mathbf{A}_T)^{-1} \mathbf{A}_T^\mathrm{H}(\mathbf{A}_T \odot \mathbf{\Omega}_q) -\mathrm{diag}(\mathbf{\Gamma}_q)=\mathbf{0}_L,$$
		for all $q=0,1,\ldots,Q-1.$
		
		%From Lemma 1, we have $\hat{\mathbf{h}}$ is closest to the channel response $\mathbf{h}_{N/2}$.
		%Therefore, the second criterion of pilot design should minimize following equation
		%\begin{equation}\label{equation_ICI}
		%E\left( \left\| \sum_{q=0}^{Q-1} (\mathbf{A}_T^H\mathbf{A}_T)^{-1} \mathbf{A}_T^H(\mathbf{A}_T \odot \mathbf{\Omega}_q) \mathbf{g}_q -\mathbf{h}_{N/2} \right\|^2 \right),
		%\end{equation}
		%where $\mathbf{h}_{N/2}$ is the channel response at $N/2$.
		Note that an efficient pilot sequence should work in a variety of channels. Hence, the second criterion ($\mathrm{CR}_2$) of pilot design is as follows:
		\begin{equation}\label{equation_ICI2}
			\arg \min_\mathbf{a} \left\| (\mathbf{A}_T^\mathrm{H}\mathbf{A}_T)^{-1} \mathbf{A}_T^\mathrm{H}(\mathbf{A}_T \odot \mathbf{\Omega}_q) -\mathrm{diag}(\mathbf{\Gamma}_q) \right\|_F^2,
		\end{equation}
		where $\mathbf{\Gamma}_q$ is given in (\ref{equation_Gamma}) and $\omega_q^{(l)}\in [-F_r, F_r]$.

To simplify the optimization algorithm, we replace $\mathrm{CR}_1$ and $\mathrm{CR}_2$ comprehensively with ISL of O-AF as follows:

		\begin{equation}\label{Opt_criterion2}
			\arg \min_\mathbf{a}
			\sum_{\substack{\tau=-Z\\ \tau\neq0}}^{Z}\int_{-F_r}^{F_r} |\widetilde{AF}_{\mathbf{a}}(\tau,f)|^2\mathrm{d} f,
		\end{equation}
  Let us call this $\mathrm{CR}_3$. A detail discussion about the relation between $\mathrm{CR}_1$, $\mathrm{CR}_2$, and $\mathrm{CR}_3$ is given in Appendix I.

Thus a sequence needs to have a low ambiguity sidelobe within the region $[-Z,Z]\times [-F_r,F_r]$, where $f\in [-F_r,F_r]$ is a continuous variable and hence can take fractional values, to be used as a pilot sequence for optimal channel estimation in DSC. The sequences which satisfy this condition are termed as O-LAZ sequences in this paper. In the next section we will derive some properties of O-LAZ sequences and also propose a construction using a modified ITROX algorithm.
		
% \blue{By Eq. (\ref{Opt_criterion2}), it is easy to see that the pilot criterion CR3 is exactly the ISL of O-AF. Therefore, O-LAZ sequences with an O-ZAZ size of $Z \times F_r$ can serve as optimal pilot sequences for use in DSC channel estimation.}
%\blue{From Eq. (\ref{Opt_criterion2}), we can see that O-AF is closely related to channel estimation in DSC. We will discuss how to design such sequences in Section IV.}

\section{O-LAZ Sequence: Properties and Construction}

To design O-LAZ sequences, which are required for optimal channel estimation in DSC, we introduce a new metric called oversampled ambiguity function (O-AF), as traditional AF considers only integer Doppler shifts. Before proceeding further, we formally define O-AF as follows.

		\subsection{Oversampled Ambiguity Function (O-AF)}
		
		\begin{Definition}
			For two length $N$ complex-valued sequences $\mathbf{a}$, $\mathbf{b}$, let $\widetilde{AF}_{\mathbf{a},\mathbf{b}}(\tau,f)$ denote the oversampled periodic cross-ambiguity function (O-PCAF) between $\mathbf{a}$ and $\mathbf{b}$ at time-shift $\tau$ and frequency-shift $f$. The $\widetilde{AF}_{\mathbf{a},\mathbf{b}}(\tau,f)$ is defined by
			\begin{equation}\label{equation_OS_PCAF}
				\widetilde{AF}_{\mathbf{a},\mathbf{b}}(\tau,f)= \sum_{k=0}^{N-1} a[k]b^*[\lfloor k+\tau\rfloor_N]\zeta_N^{fk} =\langle\mathbf{a}\odot\mathbf{f},S^{-\tau}(\mathbf{b})\rangle,
			\end{equation}
			where $\mathbf{f}=[1,\zeta_N^{f},\ldots,\zeta_N^{f(N-1)}]^\mathrm{T},\tau\in\mathbb{Z},f\in \mathbb{R}$.
			In particular, when $\mathbf{a}=\mathbf{b}$, $\widetilde{AF}_{\mathbf{a},\mathbf{b}}(\tau,f)$ is written as $\widetilde{AF}_{\mathbf{a}}(\tau,f)$ and called the oversampled periodic auto-ambiguity function (O-PAAF) of $\mathbf{a}$ at time-shift $\tau$ and frequency-shift $f$.
		\end{Definition}
		
		\begin{Remark}
             Since the Doppler shift $f$ is a continuous variable and delay shift $\tau$ is an integer in \eqref{Opt_criterion2}, we only considered oversampling the Doppler axis while defining O-AF. This helps us to accurately measure the channel estimation performance of the pilot sequence, under DSC.
		\end{Remark}
		
		\begin{Remark}
			As a generalization of AF, the O-AF has some special properties.
			\begin{itemize}
				\item When $f$ in $\widetilde{AF}_\mathbf{a}(\tau,f)$ is taken as an integer, O-AF becomes traditional AF, i.e., $\widetilde{AF}_\mathbf{a}(\tau,f)=AF_\mathbf{a}(\tau,f)$ if $f\in \mathbb{Z}$.
				\item $\widetilde{AF}_\mathbf{a}(\tau,f)$ is a periodic function with respect to $\tau$, i.e., $\widetilde{AF}_\mathbf{a}(\tau+kN,f)=\widetilde{AF}_\mathbf{a}(\tau,f)$ for any integer $k$.
				%\item $AF_\mathbf{a}(\tau,f)$ is symmetric about the origin, however $\widetilde{AF}_\mathbf{a}(\tau,f)$ is not, i.e., $|AF(-\tau,-f)|=|AF(\tau,f)|$, $|\widetilde{AF}_\mathbf{a}(-\tau,-f)| \neq |\widetilde{AF}_\mathbf{a}(\tau,f)|$.
				\item $AF_\mathbf{a}(\tau,f)$ and $\widetilde{AF}_\mathbf{a}(\tau,f)$ both have the constant volume property, i.e., $\sum_{\tau=0}^{N-1}\sum_{f=0}^{N-1}|AF_\mathbf{a}(\tau,f)|^2=N^3$, \\ $\sum_{\tau=0}^{N-1}[\int_{0}^{N}|\widetilde{AF}_\mathbf{a}(\tau,f)|^2df]=N^3$.
				\item For any unimodular sequence $\mathbf{a}$, its O-AF can be determined on zero delay axis, i.e., $|\widetilde{AF}_\mathbf{a}(0,f)|= \frac{\sin (f\pi)}{\sin (f\pi/N)}$.
			\end{itemize}
		\end{Remark}
		
		Sequences with zero/low oversampled ambiguity zone are very important for practical applications.
		In order to characterize the values of O-AF within the target zone, let us introduce the definition of the ISL of O-AF similar to (\ref{equation_ZAZ}).
		\begin{Definition}
			Let $\mathbf{a}=[a[0],a[1],\ldots,a[N-1]]^\mathrm{T}$ be a sequence of length $N$, the ISL of O-AF is define by
			\begin{equation}\label{equeation_OS_ISL}
				\mathrm{ISL}_{\widetilde{AF}}(\mathbf{a};Z\times F_r)=\sum_{\substack{\tau=-Z\\ \tau\neq0}}^{Z}\int_{f=-F_r}^{F_r} |\widetilde{AF}_{\mathbf{a}}(\tau,f)|^2\mathrm{d} f,
			\end{equation}
			where $Z\times Fr$ is the size of oversampled ambiguity zone.
		\end{Definition}
		
		Similarly, we can define oversampled zero/low ambiguity zone (O-ZAZ/O-LAZ) sequence.
		\begin{Definition}
			A sequence $\mathbf{a}$ of length $N$ is called a oversampled zero ambiguity zone (O-ZAZ) sequence with O-ZAZ size $Z\times F_r$, if
			\begin{equation}\label{equation_ZOS_AZ}
				\mathrm{ISL}_{\widetilde{AF}}(\mathbf{a};Z\times F_r)=0.
			\end{equation}
			Similarly, $\mathbf{a}$ is called a oversampled low ambiguity zone (O-LAZ) sequence with low O-LAZ size $Z\times F_r$, if
			\begin{equation}\label{equation_LOS_AZ}
				\mathrm{ISL}_{\widetilde{AF}}(\mathbf{a};Z\times F_r)<\varepsilon,
			\end{equation}
   where $\varepsilon>0$ is a small constant.
		\end{Definition}
		
In fast time-varying channel estimation, it is usually necessary to have $Z$ greater than the maximum normalized multipath delay and $F_r$ greater than the maximum normalized Doppler frequency shift (Detailed calculations are given in Section III).
		For example, let us consider the 5G scenario and the Extended Vehicular A Model (EVA) channel defined in 3GPP \cite{conformance20113rd}, assuming the number of subcarriers $N=128$, the carrier frequency $f_c=5.4 \mathrm{GHz}$, the interval of subcarriers $\Delta f=15\mathrm{KHz}$, maximum multipath delay $\tau_{\max}=2.5100 \mathrm{\mu s}$, maximum relative speed $v_{\max}=600\mathrm{Km/h}$. Then O-LAZ size need to satisfy $Z\geq 5, F_r\geq 0.2$ (Detailed calculations are given in Section V).
		\subsection{O-LAZ Sequence Design Based on ITROX}
		{In this subsection, we propose a new algorithm, OA-ITROX, based on the original ITROX algorithm \cite{soltanalian2012computational}. New O-LAZ sequences are obtained by minimizing the ISL metric, defined in (\ref{equeation_OS_ISL}). The optimization problem can be written as follows:}
		\begin{align}\label{Opt_OS_ISL}
			& \arg \min_\mathbf{a} \mathrm{ISL}_{\widetilde{AF}}(\mathbf{a};Z\times F_r), \\
			&\hbox{subject to }|a_k|=1,k=0,1,\ldots,N-1. \nonumber
		\end{align}
		
		Since (\ref{equeation_OS_ISL}) contains an integral operation, it is challenging to directly optimize $\mathrm{ISL}_{\widetilde{AF}}(\mathbf{a};Z\times F_r)$.
		Since $\widetilde{AF}_\mathbf{a}(\tau,f)$ can be well approximated when $f$ is very small, we can \textcolor{black}{verify} that (\ref{Opt_OS_ISL}) is equivalent to the following simpler problem:
		\begin{align}\label{Opt_OS_MISL}
			& \arg \min_\mathbf{a} \sum_{\substack{\tau=-Z\\ \tau\neq0}}^{Z}  f_\delta \sum_{f\in\mathcal{F}} |\widetilde{AF}_{\mathbf{a}}(\tau,f)|^2, \\
			&\hbox{subject to }|a_k|=1,k=0,1,\ldots,N-1, \nonumber
		\end{align}
		where $\mathcal{F}=\{0,\pm f_\delta,\pm 2f_\delta,\ldots,\pm F_r\}$ (Without loss of generality, we can assume that $F_r$ is a multiple of $f_\delta$ and $F_r=(M-1)f_\delta$).
		
		Before introducing the OA-ITROX algorithm, let us represent the O-PAAF of sequences using matrices.
		We first define the periodic diagonal elements. The elements of the $k$-th periodic diagonal of matrix $\mathbf{A}$ are given by
		$$
		\mathrm{diag}(\mathbf{A}, k)=[[\mathbf{A}]_{0,k},\ldots,[\mathbf{A}]_{i,\lfloor k+i\rfloor_N},\ldots, [\mathbf{A}]_{N-1,k-1}]^\mathrm{T},
		$$
		where $k=0,1,\ldots,N-1$. For example, let $\mathbf{A}$ be a matrix of size $4\times 4$.
		Then we have
		\begin{align*}
			\mathrm{diag}(\mathbf{A}, 0)&=[A_{0,0},A_{1,1},A_{2,2},A_{3,3}]^\mathrm{T},\\
			\mathrm{diag}(\mathbf{A}, 1)&=[A_{0,1},A_{1,2},A_{2,3},A_{3,0}]^\mathrm{T},\\
			\mathrm{diag}(\mathbf{A}, 2)&=[A_{0,2},A_{1,3},A_{2,0},A_{3,1}]^\mathrm{T},\\
			\mathrm{diag}(\mathbf{A}, 3)&=[A_{0,3},A_{1,0},A_{2,1},A_{3,1}]^\mathrm{T}.\\
		\end{align*}
		Let $\mathrm{DS}(\mathbf{A}, k)$ represents the sum of all the elements of the $k$-th periodic diagonal of matrix $\mathbf{A}$, i.e.,
		\begin{equation}\label{equation_DS}
			\mathrm{DS}(\mathbf{A}, k)=\sum \mathrm{diag}(\mathbf{A}, k).
		\end{equation}
		By (\ref{equation_DS}), we can calculate the O-PAAF magnitudes by the following equation:
		{
			\begin{equation}\label{equation_DS_OS}
				|\widetilde{AF}_{\mathbf{a}}(\tau,f)|=
				\left\{
				\begin{array}{ll}
					|\mathrm{DS}((\mathbf{a}\odot\mathbf{f})\mathbf{a}^\mathrm{H},\tau)|, & f\geq0; \\
					|\mathrm{DS}(\mathbf{a}(\mathbf{a}\oslash \mathbf{f})^\mathrm{H},N-\tau)|, & f<0,
				\end{array}
				\right.
			\end{equation}
		}
		where $\mathbf{f}=[1,\zeta_N^{f},\ldots,\zeta_N^{f(N-1)}]^\mathrm{T}$, $\tau=0,1,\ldots,N-1$, and $f\in \mathbb{R}$.
		
		Furthermore, we can obtain the complete matrix representation of the O-PAAF of sequences in (\ref{equation_mat_OS}),
		where \\$\mathbf{f}_m=[1,\zeta_N^{mf_\delta},\ldots, \zeta_N^{mf_\delta(N-1)}]^\mathrm{T},m=0,1,\ldots,M-1$, $\mathbf{A}_{0,0}=\mathbf{a}\mathbf{a}^\mathrm{H}$ and
		$\mathbf{A}_{i,j}=(\mathbf{a}\odot\mathbf{f}_i)(\mathbf{a}\oslash \mathbf{f}_j)^\mathrm{H}$.
	%	\begin{figure*}[ht]
			\begin{equation}\label{equation_mat_OS}
				\begin{split}
				&	\left[
					\begin{array}{c}
						\mathbf{a} \\
						\mathbf{a}\odot\mathbf{f}_1 \\
						\vdots \\
						\mathbf{a}\odot\mathbf{f}_{M-1} \\
					\end{array}
					\right]
					\left[
					\begin{array}{cccc}
						\mathbf{a}^\mathrm{H} & (\mathbf{a}\oslash \mathbf{f}_1)^\mathrm{H} & \cdots & (\mathbf{a}\oslash \mathbf{f}_{M-1})^\mathrm{H} \\
					\end{array}
					\right]
					 =
					\left[
					\begin{array}{cccc}
						\mathbf{A}_{0,0} & \mathbf{A}_{0,1} & \cdots & \mathbf{A}_{0,M-1} \\
						\mathbf{A}_{1,0} & \mathbf{A}_{1,1} & \cdots & \mathbf{A}_{1,M-1} \\
						\vdots & \vdots &  & \vdots \\
						\mathbf{A}_{M-1,0} & \mathbf{A}_{M-1,1} & \cdots & \mathbf{A}_{M-1,M-1} \\
					\end{array}
					\right].
				\end{split}
			\end{equation}
%		\end{figure*}
		According to (\ref{equation_DS_OS}), the O-PAAF matrix representation can be divided into three case.\\
		{
			%\begin{itemize}
				Case I:
				 $|\widetilde{AF}_{\mathbf{a}}(\tau,0)|=|\mathrm{DS}(\mathbf{A}_{0,0},\tau)|;$\\
				Case II: $|\widetilde{AF}_{\mathbf{a}}(\tau,m f_\delta)| =|\mathrm{DS}(\mathbf{A}_{m,0},\tau)|;$\\
				Case III: $|\widetilde{AF}_{\mathbf{a}}(\tau,-m f_\delta)| =|\mathrm{DS}(\mathbf{A}_{0,m},N-\tau)|.$
			%\end{itemize}
		}
		
		Let
		\begin{equation}\label{}
			\Lambda=\{\mathbf{X}|\mathbf{X}=\mathbf{x}\mathbf{y}^\mathrm{H} \},
		\end{equation}
		where both $\mathbf{x,y}$ are unimodular sequences of length $MN$.
		%Given a unimodular sequence $\mathbf{a}$, let us construct $\mathbf{X}_\mathbf{a}$ as follows:
		In OA-ITROX algorithm, $\mathbf{X}$ is always generated by the unimodular sequence $\mathbf{a}$ which is being optimized, so it is commonly denoted as $\mathbf{X}_\mathbf{a}$.
		$\mathbf{X}_\mathbf{a}$ is given by
		\begin{equation}\label{equation_X_a}
			\mathbf{X}_\mathbf{a}=\mathbf{x}_\mathbf{a}\mathbf{y}^\mathrm{H}_\mathbf{a},
		\end{equation}
		where
		$$
		\mathbf{x}_\mathbf{a}=
		\left[
		\begin{array}{c}
			\mathbf{a} \\
			\mathbf{a}\odot\mathbf{f}_1 \\
			\vdots \\
			\mathbf{a}\odot\mathbf{f}_{M-1} \\
		\end{array}
		\right],
		\mathbf{y}_\mathbf{a}=
		\left[
		\begin{array}{c}
			\mathbf{a} \\
			\mathbf{a}\oslash\mathbf{f}_1 \\
			\vdots \\
			\mathbf{a}\oslash\mathbf{f}_{M-1} \\
		\end{array}
		\right].
		$$
		Also, let
		\begin{equation}\label{}
			\Gamma_Z=\left\{\mathbf{Y}|\mathbf{Y}=
			\left[
			\begin{matrix}
				\mathbf{Y}_{0,0} & \cdots & \mathbf{Y}_{0,M-1} \\
				\vdots &  & \vdots \\
				\mathbf{Y}_{M-1,0} & \cdots & \mathbf{Y}_{M-1,M-1} \\
			\end{matrix}
			\right]
			\right\},
		\end{equation}
		where $\mathbf{Y}$ is a matrix of size $MN\times MN$, containing block matrices $\mathbf{Y}_{m,n}$ each of size $N\times N$, where $\mathbf{Y}_{0,0}$ satisfies
		{
			$$
			\mathrm{DS}(\mathbf{Y}_{0,0},\tau)=
			\left\{
			\begin{array}{ll}
				N, & \tau=0; \\
				0, & \tau=1,\ldots,Z,N-Z,\ldots,N-1,
			\end{array}
			\right.
			$$
			$\mathbf{Y}_{m,0}$, $\mathbf{Y}_{0,m}$ satisfy
			\begin{align*}
				& \mathrm{DS}(\mathbf{Y}_{m,0},\tau)
				= \mathrm{DS}(\mathbf{Y}_{0,m},N-\tau) \\
				=&\left\{
				\begin{array}{ll}
					\sum_{n=0}^{N-1}\zeta_N^{mf_\delta n}, & \tau=0; \\
					0, & \tau=1,\ldots,Z,N-Z,\ldots,N-1.
				\end{array}
				\right.
			\end{align*}
		}
		
		The proposed OA-ITROX is a cyclic iterative algorithm that constantly finds the optimal projection (for the matrix Frobenius norm) between $\Lambda$ and $\Gamma_Z$.
		In the following theorem, we study the orthogonal projection of an element of $\Lambda$ on $\Gamma_Z$. This process can be described as follows:
		\begin{align}\label{}
			\arg\min_\mathbf{Y} &~~ \|\mathbf{Y}-\mathbf{X}\|_F^2, \\
			\hbox{s.t.~~}&\mathbf{Y}\in\Gamma_Z. \notag
		\end{align}
		
		\begin{Theorem}\label{theorem ITROX 1}
			Let $\mathbf{Y}=\mathbf{X}^\perp$ be the optimal projection of $\mathbf{X}\in \Lambda$ on $\Gamma_Z$, then $\mathbf{Y}$ can be constructed through the following cases:
			
			Case I: For {$\tau=0,Z+1,Z+2,\ldots,N-1-Z$},
			$$
			\mathrm{diag}(\mathbf{Y}_{m,0},\tau)=\mathrm{diag}(\mathbf{X}_{m,0},\tau),
			$$
			$$
			\mathrm{diag}(\mathbf{Y}_{0,m},\tau)=\mathrm{diag}(\mathbf{X}_{0,m},\tau),
			$$
			where $m=0,1,\ldots,M-1.$
			
			Case II: For {$\tau=1,\ldots,Z,N-Z,\ldots,N-1$},
			{
				$$
				\mathrm{diag}(\mathbf{Y}_{m,0},\tau)=\mathrm{diag}(\mathbf{X}_{m,0},\tau) -\frac{1}{N}\mathrm{DS}(\mathbf{X}_{m,0},\tau),
				$$
				$$\mathrm{diag}(\mathbf{Y}_{0,m},\tau)=\mathrm{diag}(\mathbf{X}_{0,m},\tau) -\frac{1}{N}\mathrm{DS}(\mathbf{X}_{0,m},\tau),
				$$
			}
			where $m=0,1,\ldots,M-1.$
			
			Case III: For $m\neq 0$ or $n\neq 0$,
			$$
			\mathbf{Y}_{m,n}=\mathbf{X}_{m,n},
			$$
			where $\mathbf{X}_{m,n}=(\mathbf{a}\odot\mathbf{f}_m)(\mathbf{a}\oslash\mathbf{f}_n)^\mathrm{H}, m,n=0,1,\ldots,M-1$.
		\end{Theorem}
		
		\begin{IEEEproof}
			We have
			\begin{equation*}
				\|\mathbf{Y}-\mathbf{X}\|_F^2= \sum_{\substack{\tau=-Z \\ \tau\neq0}}^{Z}\sum_{m=0}^{M-1} \big(\|\mathrm{diag}(\mathbf{Y}_{m,0},\tau)-\mathrm{diag}(\mathbf{X}_{m,0},\tau)\|^2 \\
				+ \|\mathrm{diag}(\mathbf{Y}_{0,m},\tau)-\mathrm{diag}(\mathbf{X}_{0,m},\tau)\|^2 \big).
			\end{equation*}
			Our goal is to minimize $\|\mathbf{Y}-\mathbf{X}\|_F^2$ for $\mathbf{Y}\in\Gamma_Z$.
			Hence, we can minimize $g_{m,\tau}^{\mathrm{col}}(\mathbf{Y}):= \|\mathrm{diag}(\mathbf{Y}_{m,0},\tau)-\mathrm{diag}(\mathbf{X}_{m,0},\tau)\|^2$ and $g_{m,\tau}^{\mathrm{row}}(\mathbf{Y}):= \|\mathrm{diag}(\mathbf{Y}_{0,m},\tau)-\mathrm{diag}(\mathbf{X}_{0,m},\tau)\|^2$ for all $m=0,1,\ldots,M-1$, $\tau=1,\ldots,Z,N-Z,\ldots,N-1$.
			Denote $x_k^{(m,0,\tau)}=[\mathrm{diag}(\mathbf{X}_{m,0},\tau)]_k$ and $y_k^{(m,0,\tau)}=[\mathrm{diag}(\mathbf{Y}_{m,0},\tau)]_k$, where $k=0,1,\ldots,N-1$.
			Then
			$$
			g_{m,\tau}^{\mathrm{col}}(\mathbf{Y})=\sum_{k=0}^{N-1} |y_k^{(m,0,\tau)}-x_k^{(m,0,\tau)}|^2,
			$$
			such that $\sum_{k=0}^{N-1}y_k^{(m,0,\tau)}=0$. Using the Cauchy-Schwarz inequality we have that
			$$
			g_{m,\tau}^{\mathrm{col}}(\mathbf{Y})\geq \frac{1}{N} \left|\sum_{k=0}^{N-1}\left(y_k^{(m,0,\tau)}-x_k^{(m,0,\tau)}\right)\right|^2.
			$$
			Note that the equality holds if and only if $y_k^{(m,0,\tau)}=x_k^{(m,0,\tau)}- \frac{1}{N} \sum_{k=0}^{N-1}x_k^{(m,0,\tau)}$.
			Returning to the matrix representation, we can infer that the minimum value of $g_{m,\tau}^{\mathrm{col}}(\mathbf{Y})$ is achieved at the point
			$$
			\mathrm{diag}(\mathbf{Y}_{m,0},\tau)=\mathrm{diag}(\mathbf{X}_{m,0},\tau) -\frac{1}{N}\mathrm{DS}(\mathbf{X}_{m,0},\tau).
			$$
			Similarly, we can prove that the minimum value of $g_{m,\tau}^{\mathrm{row}}(\mathbf{Y})$ is achieved when
			$$\mathrm{diag}(\mathbf{Y}_{0,m},\tau)=\mathrm{diag}(\mathbf{X}_{0,m},\tau) -\frac{1}{N}\mathrm{DS}(\mathbf{X}_{0,m},\tau).
			$$
		\end{IEEEproof}
		
		In the next theorem, we find the optimal projection of $\mathbf{Y}\in \Gamma_Z$ in $\Lambda$, described as follows:
		\begin{align}\label{}
			\arg\min_\mathbf{X} &~~ \|\mathbf{X}-\mathbf{Y}\|_F^2, \\
			\hbox{s.t.~~}&\mathbf{X}\in\Lambda. \notag
		\end{align}
		
		\begin{Theorem}\label{theorem ITROX 2}
			Let $\mathbf{X}=\mathbf{Y}^\perp$ be the optimal projection of $\mathbf{Y}\in \Gamma_Z$ on $\Lambda$.
			Suppose $\mathbf{Y}$ has the singular value decomposition (SVD)
			\begin{equation}
				\mathbf{S\Sigma V}^\mathrm{H}=\mathbf{Y},
			\end{equation}
			then $\mathbf{X}$ is given by
			\begin{equation}
				\mathbf{X}=MN\mathbf{sv}^\mathrm{H},
			\end{equation}
			where $\mathbf{s,v}$ are the left and right singular vector corresponding to the largest singular value of $\mathbf{Y}$, respectively.
		\end{Theorem}
		
		\begin{IEEEproof}
		The proof is omitted, as it is similar to the proof of the optimal rank-1 matrix approximation given in \cite{soltanalian2012computational}.
		\end{IEEEproof}
			%This can be proved using Ecart-Young's Theorem for low-rank matrix approximation.
		
		Note that we cannot construct $\mathbf{a}$ from $\mathbf{X}$ in \textit{Theorem \ref{theorem ITROX 2}}.
		Hence, the process of generating sequence $\mathbf{a}$ from matrix $\mathbf{X}$ is described through the following optimization problem.
		\begin{align}\label{equation_almost}
			\arg\min_\mathbf{a} &~~ \|\mathbf{X_a}-\mathbf{X}\|_F^2, \\
			\hbox{s.t.~~}& \mathbf{X}_\mathbf{a}=\mathbf{x}_\mathbf{a}\mathbf{y}^\mathrm{H}_\mathbf{a},\notag \\
			& \mathbf{X}=MN\mathbf{sv}^\mathrm{H}. \notag
		\end{align}
		
		It is equivalent to solving the following optimization problem:
		\begin{align}\label{Opt_3}
			\arg\min_\mathbf{a} &~~ \|\mathbf{x}_\mathbf{a}-\sqrt{MN}\mathbf{s}\|_F^2 +\|\mathbf{y}_\mathbf{a}-\sqrt{MN}\mathbf{v}\|_F^2.
		\end{align}
		
		\begin{Theorem}\label{theorem ITROX 3}
			The problem (\ref{Opt_3}) has a closed-form solution, which is given by
			\begin{equation*}
				a[k]=\mathrm{exp}\bigg\{i\varphi\bigg(\frac{1}{M}\sum_{m=0}^{M-1}\sqrt{MN}s[k+mM]/\zeta_N^{m f_\delta k} \\ +\frac{1}{M}\sum_{m=0}^{M-1}\sqrt{MN}v[k+mM]\cdot\zeta_N^{m f_\delta k}\bigg)\bigg\},
			\end{equation*}
			where $i=\sqrt{-1}$ is the imaginary unit, $k=0,1,\ldots,N-1$.
		\end{Theorem}
		
		\begin{IEEEproof}
			The proof is omitted, as it is similar to the proof of the optimal approximation in the AF-CAO algorithm  \cite{li2008signal}.
		\end{IEEEproof}

		The proposed OA-ITROX for the local minimization of the ISL metric in (\ref{Opt_OS_MISL}) can be summarized in \textbf{Algorithm 1}.
		%Fig. \ref{fig_liucheng} shows the program flow chart of \textbf{Algorithm 1}.
		\begin{table}[ht]
			\centering
			\label{Algorithm ITROX}
			\begin{tabular}{p{17cm}}
				\hline\hline
				\textbf{Algorithm 1:} The OA-ITROX algorithm \\ \hline
				\textbf{Input:} Sequence length $N$; O-LAZ size $Z\times F_r$; Maximum iterations number $\mathrm{MaxItrNum}$;
				Initial point $\mathbf{a}^{0}$ is a random sequence. Let $t=1$. \\
				\textbf{Step 1:} Calculate $\mathbf{X}_\mathbf{a}^{t-1}= ({\mathbf{x}}_\mathbf{a}^{t-1})({\mathbf{y}}_\mathbf{a}^{t-1})^\mathrm{H}$ and find $\mathbf{Y}^t\in\Gamma_Z$ according to \textit{Theorem \ref{theorem ITROX 1}}. \\
				\textbf{Step 2:} Compute the SVD of $\mathbf{Y}^t$ and find $\mathbf{X}^t \in\Lambda$ according to \textit{Theorem \ref{theorem ITROX 2}}. \\
				\textbf{Step 3:} Let the left-singular vector and right-singular vector of corresponding to the maximum singular value of $\mathbf{X}^t$ be $\mathbf{s,v}$, respectively. Then the unimodular sequences ${\mathbf{a}}^t$ is obtained by \textit{Theorem \ref{theorem ITROX 3}}.\\
				\textbf{Iteration:} Repeat Steps 1-3 until some stop criterion is satisfied, e.g., $\|\mathbf{a}^t-\mathbf{a}^{t-1}\|^2\leq \varepsilon$, where $\varepsilon$ is a predefined threshold, or $t\leq\mathrm{MaxItrNum}$. Otherwise, update $t=t+1$ and continue the iterations. \\
				\textbf{Output:} Set the optimized sequence as $\mathbf{a}$. \\
				\hline\hline
			\end{tabular}
		\end{table}
		
%		\begin{figure}[ht]
%			\centering
%			\includegraphics[width=7.481cm]{fig/liuchengtu}
%			\caption{The program flow chart of \textbf{Algorithm 1}.}\label{fig_liucheng}
%		\end{figure}
		
		\begin{Remark}
		In the proposed OA-IROTX algorithm, the choice of the parameter $f_\delta$ is important. From the principles of Riemann integrals in mathematical analysis \cite{Zorich2015}, it can be observed that if $f_\delta$ is too large, (\ref{Opt_OS_MISL}) cannot closely approximate (\ref{Opt_OS_ISL}). However, if $f_\delta$ is too small, the complexity of the algorithm will increase significantly.
Taking into account the two reasons mentioned above, we suggest setting the value of $f_\delta$ between 0.1 and 0.2. If the parameter $f_\delta$ is set to 1, then the objective function reduces to the traditional AF $AF_\mathbf{a}(\tau,f)$.
		\end{Remark}
		\begin{Remark}\label{rem5}
			The primary computational cost of \textbf{Algorithm 1} comes from Step 2. Therefore, the complexity of \textbf{Algorithm 1} is determined by the complexity of SVD decomposition. In general, for matrix $A$ of size $m\times n$, the complexity of SVD is $\mathcal{O}(mn\times \min(m,n))$ \cite{golub2013matrix}. Since we are concerned with the maximum singular value only, we can employ power method to reduce the complexity from $\mathcal{O}(mn\times \min(m,n))$ to $\mathcal{O}(mn)$ \cite{golub2013matrix}. Hence, in this work, the complexity of OA-ITROX is $\mathcal{O}\left(\left(\frac{NF_r}{f_\delta}\right)^2\right)$.
		\end{Remark}

\begin{Remark}\label{r6}
ITROX outperforms some other algorithms from the following perspectives:
\begin{itemize}
    \item AF-CAO requires the correlation matrix to be a block diagonal matrix.
    \item AISO requires the first majorization function to be a real-valued function.
    \item GD has excessively high algorithmic complexity and hence is not suitable for constructing O-LAZ sequences.
\end{itemize}

\end{Remark}

		\section{Numerical Experiments}
%		In this section, we first show the performance of the proposed OA-ITROX algorithm in solving the optimization problem (\ref{Opt_OS_MISL}). Then we conduct channel estimation simulation experiments using the O-LAZ sequences obtained from the OA-ITROX algorithm over a DSC model, and compare their performance with traditional sequences like ZC sequences and $m$-sequences.
		
		\subsection{Performance Analysis of OA-ITROX Algorithm}
		Using the proposed algorithm, we have designed an unimodular sequence of length $N = 128$ with low O-AF zone $(Z\times F_r)=(32\times 0.2)$.
		%\footnote{All the optimized sequences in this subsection can be downloaded from the author's GitHub website: xxx.}.
		The OA-ITROX algorithm is initialized with random phase sequence.
		The maximum iteration number is set to {$2\times10^6$ and $f_\delta=0.2$.}
%		Fig. \ref{fig_OS_AF1} shows the three-dimensional figure of the O-AF of the obtained sequence generated by OA-ITROX. It can be observed that the $|\widetilde{AF}_{\mathbf{a}}(\tau,f)|$ exhibits a clear O-LAZ for $\tau \in [-Z,Z]$.
%		\begin{figure}[ht]
%			\centering
%			\includegraphics[width=8cm]{fig/fig_OS_AF1}
%			\caption{The three-dimensional figure of the O-AF of the sequence generated by OA-ITROX.}\label{fig_OS_AF1}
%		\end{figure}
		
		To provide a clear description of the performance of OA-ITROX algorithm within the O-LAZ, we define the ambiguity level (dB) as follows
		$$\mathrm{ambiguity~level}= 20\log_{10}\frac{|\widetilde{AF}_{\mathbf{a}}(\tau,f)|}{|\widetilde{AF}_{\mathbf{a}}(0,0)|}.$$
		
\begin{figure}[ht]
			\centering
			\includegraphics[width=8cm]{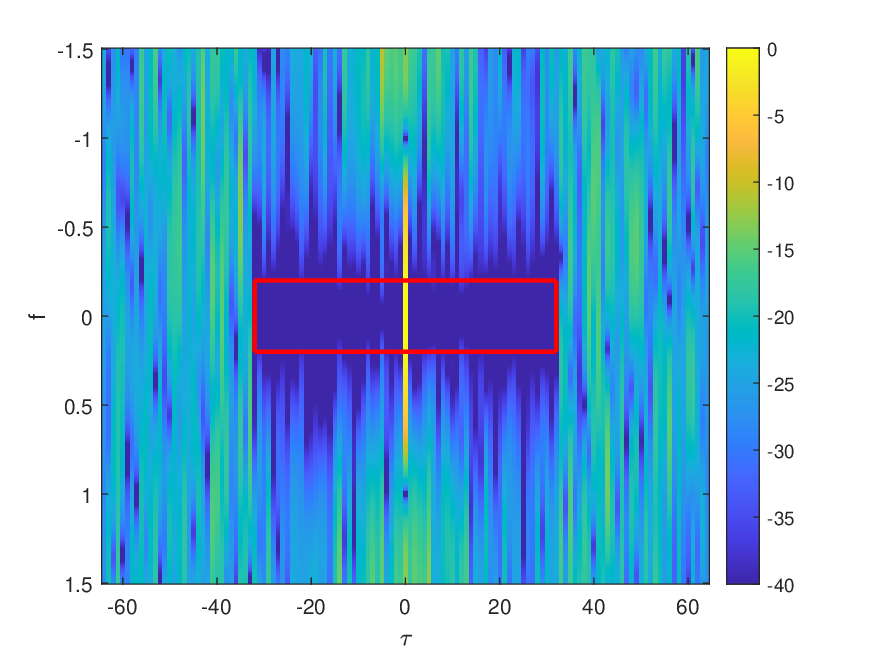}
			\caption{The planform figure of the O-AF of the o-LAZ sequence generated by OA-ITROX algorithm.}\label{fig_OS_AF2}
		\end{figure}

		The planform figure of the O-AF of the generated sequence is shown in Fig. \ref{fig_OS_AF2}. The the O-LAZ is highlighted with a red solid line box.
		It can be observed that within the O-LAZ, the ambiguity level of O-AF is consistently below -40 dB.
		
\begin{figure}[ht]
			\centering
			\includegraphics[width=8cm]{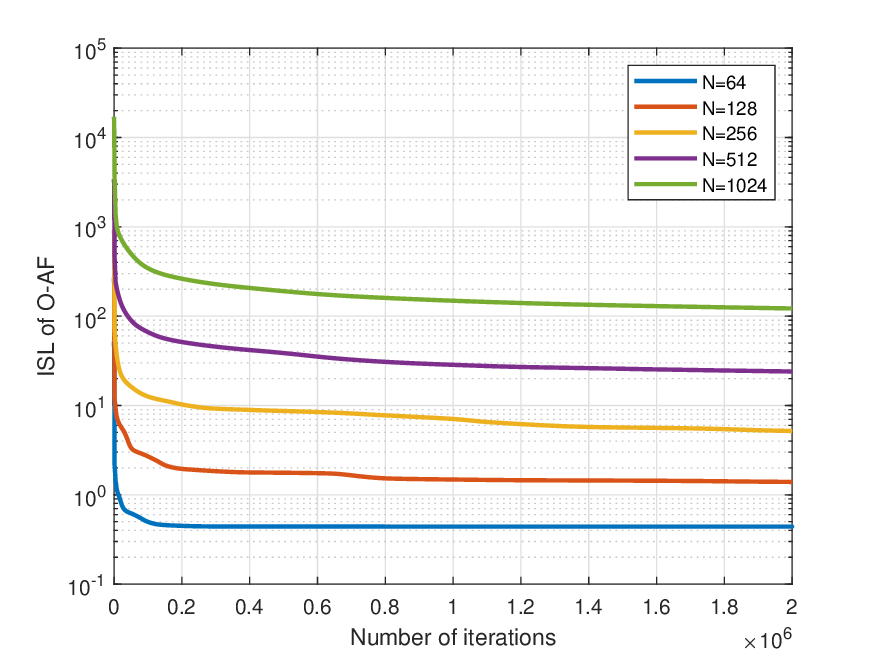}
			\caption{Semi-log plot of the ISL with respect to the iteration index.}\label{fig_OS_AF_ISL}
		\end{figure}

		Fig. \ref{fig_OS_AF_ISL} shows the ISL value of the objective function with the number of iterations during the execution of the OA-ITROX algorithm.
		{In this experiment, the sequence lengths and O-LAZ size are set as $N=64,128,256,512, 1024, ~Z=N/4, F_r=0.2$. This indicates that the proposed modified ITROX algorithm has good convergence in every sequence length.}
		
%		To demonstrate the robustness of the OA-ITROX algorithm across various initial points, we have fixed the parameters and run the OA-ITROX algorithm for 150 times, each time taking random initial seed sequence. For the experiment shown in Fig. \ref{fig_box}, we have considered the sequence length of $N=128$, $f_\delta=0.2$, $F_r=0.2$, maximum number of iteration $10^5$ and O-LAZ width $Z\times F_r$. We have run the OA-ITROX algorithm for $Z=4,~8,~12,~16,~20,~24,~28,$ and $32$. For each $Z$, we run the OA-ITROX algoritm 150 times, each time using a random seed sequence. In Fig. \ref{fig_box}, the red line within the box for each $Z$ shows the median of the ISL, the bottom and top of each box denotes the 25th and 75th percentile of the 150 ISLs obtained for running the OA-ITROX algorithm 150 times. Through this experiment we demonstrate that the proposed OA-ITROX algorithm is robust to initial random seed sequences and can efficiently optimize the O-AF for any given value of $Z$.
%
%		\begin{figure}[ht]
%			\centering
%			\includegraphics[width=8cm]{fig_box.pdf}
%			\caption{ISL of O-AF versus O-LAZ width $Z\times F_r$.}\label{fig_box}
%		\end{figure}

		\subsection{Performance Comparison with Traditional AF}
		As mentioned in \textit{Remark 3}, if we set $f_\delta=1$, OA-ITROX can optimize the traditional AF {which leads to traditional LAZ sequences}. In this section, we evaluate the performances of the {traditional LAZ sequences constructed using the proposed} OA-ITROX algorithm and the AF-CAO algorithm \cite{li2008signal, he2012waveform} when optimizing the traditional AF.
		
\begin{figure}
			\centering
			\subfigure[AF of sequences generated by AF-CAO \cite{li2008signal}]{
				\includegraphics[width=5.5cm]{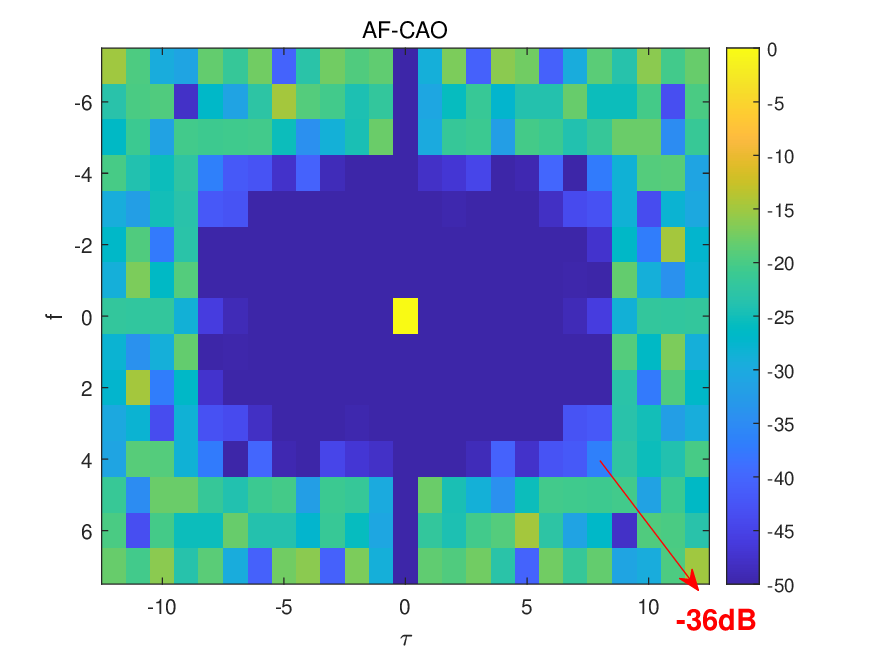}
			}
			\quad
			\subfigure[AF of sequences generated by OA-ITROX algorithm]{
				\includegraphics[width=5.5cm]{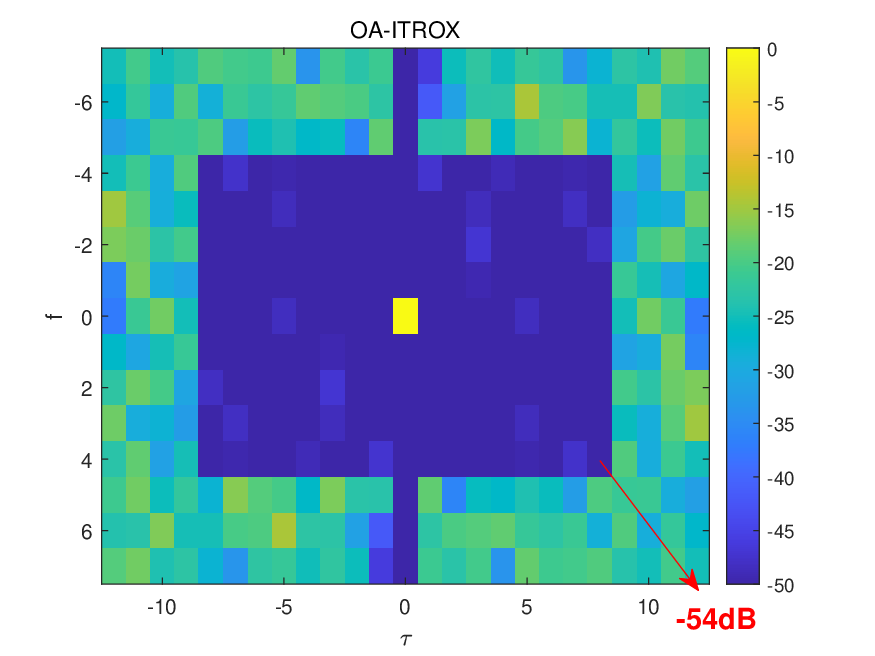}
			}
			% difference in this line
			\caption{The planform figure of the traditional AF of the sequences.}\label{fig_CAO_ITROX}
		\end{figure}

\begin{figure}[ht]
			\centering
			\includegraphics[width=8cm]{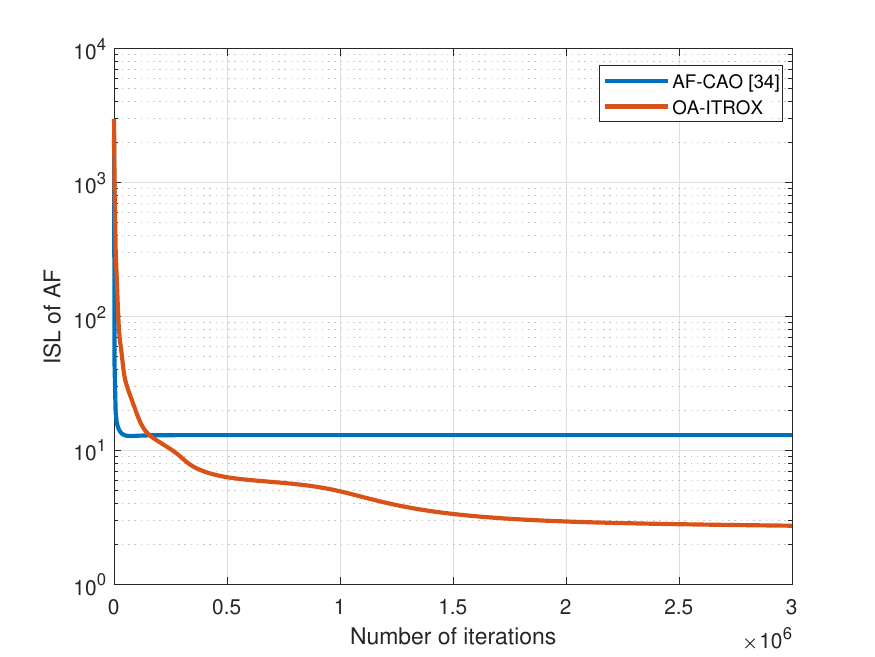}
			\caption{Semi-log plot of the ISL with respect to the iteration index.}\label{fig_CAO_ITROX_ISL}
		\end{figure}

		We consider the sequence length of $N=128$, the LAZ size of $(Z\times F_r)=(8\times4)$, the maximum iteration number of $3\times 10^6$, and a random sequence as the initial solution. Fig. \ref{fig_CAO_ITROX} shows the planform figure of {the traditional LAZ sequence constructed using} AF-CAO algorithm and OA-ITROX, respectively. {Fig. \ref{fig_CAO_ITROX_ISL} provides a comparison of ISL changes with the number of iterations.}
		Fig. \ref{fig_CAO_ITROX} and Fig. \ref{fig_CAO_ITROX_ISL} show that OA-ITROX algorithm can be employed to design sequences having good traditional AF magnitudes. Furthermore, OA-ITROX outperforms the AF-CAO algorithm in terms of AF sidelobe. Within the LAZ, sequences generated by the AF-CAO algorithm exhibit uneven sidelobes, with higher values around the four corners. In contrast, sequences generated by the OA-ITROX algorithm demonstrate uniform sidelobes within the LAZ, all about -50 dB.
		
		%	\begin{figure}[ht]
			%		\centering
			%		\includegraphics[width=9cm]{fig_CAO_ITROX}
			%		\caption{The planform figure of the traditional ambiguity function of the sequences generated by AF-CAO and OA-ITROX.}\label{fig_CAO_ITROX}
			%	\end{figure}
		%	
		
		\subsection{Simulation Results}
		%In this subsection, we evaluate the pilot sequences generated by OA-ITROX.
		%\footnote{The MATLAB codes used in our simulations can be downloaded from the author's GitHub website: xxx.}.
		We assume the number of subcarriers  $N=128$, cyclic prefix (CP) length {$N_\mathrm{CP}=32$, the IF frequency $f_c=3.4  ~\mathrm{GHz}$,} the interval of subcarriers $\Delta f=15~\mathrm{KHz}$, maximum relative speed {$v_{\max}=500~\mathrm{Km/h}$.}
		In addition, we set the power delay profile according to the EVA channel parameters defined in the 3GPP standards \cite{3gpp2019study}.
		%Please note that due to the time resolution of the OFDM system being $\frac{1}{\Delta f N}$, we present the normalized power delay profile parameters after multipath merging in the third and fourth columns of Table \ref{table_channel}, which depends on the time resolution.
		From \cite{hlawatsch2011wireless}, the maximum Doppler frequency $f_d$ is calculated as {$vf_c/c=1.57$ KHz. Consequently, the normalized Doppler frequency is 0.105.}
%		\begin{table}[ht]
%			\centering
%			\caption{EVA Delay Profile}\label{table_channel}
%			\begin{tabular}{|c|c|}
%				\hline
%				% after \\: \hline or \cline{col1-col2} \cline{col3-col4} ...
%				Tap delay (ns) & \makecell{Relative \\power (dB)}  \\ \hline
%				0 & 0 \\
%				30 & -1.5 \\
%				150 & -1.4  \\
%				310 & -3.6  \\
%				370 & -0.6  \\
%				710 & -9.1  \\
%				1090 & -7.0  \\
%				1730 & -12.0 \\
%				2510 & -16.9  \\
%				\hline
%			\end{tabular}
%		\end{table}
		
		We have designed two simulation experiments. The first simulation experiment focuses on the channel estimation performance of a single OFDM pilot symbol. As described in Subsection \ref{sec2b}, a single OFDM pilot symbol provides an estimate $\hat{\mathbf{h}}$ of the channel response $\mathbf{h}_\mathrm{mid}$. Therefore, the MSE of the midpoint instant is defined as follows:
		$$
		\mathrm{MSE}_\mathrm{mid}=E\left(\|\mathbf{h}_\mathrm{mid}-\hat{\mathbf{h}}\|_2^2\right).
		$$
		
		%{ZC sequences and $m$-sequences are two well-known sequences, which are referred to as demodulatin reference signal (DMRS) and primiary synchronization signal (PSS) in the 3GPP standards, respectively.}
		For benchmarking, we select the following three sequences as pilot sequences and conduct channel estimation simulation experiments.
		The first sequence is a Zadoff-Chu (ZC) sequence of length $N=128$, which is given by
		$$
		\mathbf{s}_\mathrm{zc}^{(\mu)}=[s_\mathrm{zc}^{(\mu)}[0], s_\mathrm{zc}^{(\mu)}[1], \ldots, s_\mathrm{zc}^{(\mu)}[127]]^\mathrm{T},
		$$
		where $s_\mathrm{zc}^{(\mu)}[k]=\zeta_{2N}^{-\mu k^2}$ and $\gcd(\mu,N)=1$. The second sequence is an extended $m$-sequence of length $N=128$, which is given by
		$$
		\mathbf{s}_\mathrm{m}=[s_\mathrm{m}[0],s_\mathrm{m}[1],\ldots,s_\mathrm{m}[126],1]^\mathrm{T},
		$$
		where $s_\mathrm{m}[k]=(-1)^{\mathrm{Tr}(\alpha^k)}$, $\alpha$ is the primitive element of Galois field $GF(2,7)$, and $\mathrm{Tr}(\cdot)$ is the Trace function over $GF(2,7)$.
%		The reason we do not directly use the $m$-sequence is that the length of the $m$-sequence is $2^k-1$. Therefore, we needed to append a `1' to the end of the $m$-sequence to make its overall length 128.
The third sequence is a traditional LAZ sequence, which is generated by AF-CAO algorithm \cite{li2008signal}.
			%\footnote{The Matlab code of CAO algorithm can be find in github xxx.} \cite{li2008signal}.
			We set the parameters for AF-CAO algorithm are as follows: $N=128, Z=32, F_r=1$ and maximum iteration number $2\times10^6$.
			Note that, since it does not consider the impact of O-AF, this sequence may serve as a baseline to measure the effective gain of O-LAZ sequences. The fourth sequence is O-LAZ sequence which is constructed using the proposed OA-ITROX algorithm.
		The input parameters for the algorithm are as follows: $N=128, Z=32, F_r=f_\delta=0.2$ and maximum iteration number $2\times10^6$. %The initial point is generalized ZC sequence with $u = \frac{N}{2Z+1}$.
		\begin{Remark}
		ZC sequences and $m$ sequences are not unique. Here, we calculated the ISL of O-AF for 64 ZC sequences and 18 $m$-sequences with parameters $Z=32,~F_r=0.2$. We have chosen the sequences with the highest and lowest ISL magnitude, and denoted them as ``The best ZC/$m$-sequence" and ``The worst ZC/$m$-sequence," respectively. Furthermore, the best ZC sequence is used as the initial seed sequence for the OA-ITROX algorithm, to construct the O-LAZ sequence.
		\end{Remark}

\begin{figure}[ht]
			\centering
			\includegraphics[width=8cm]{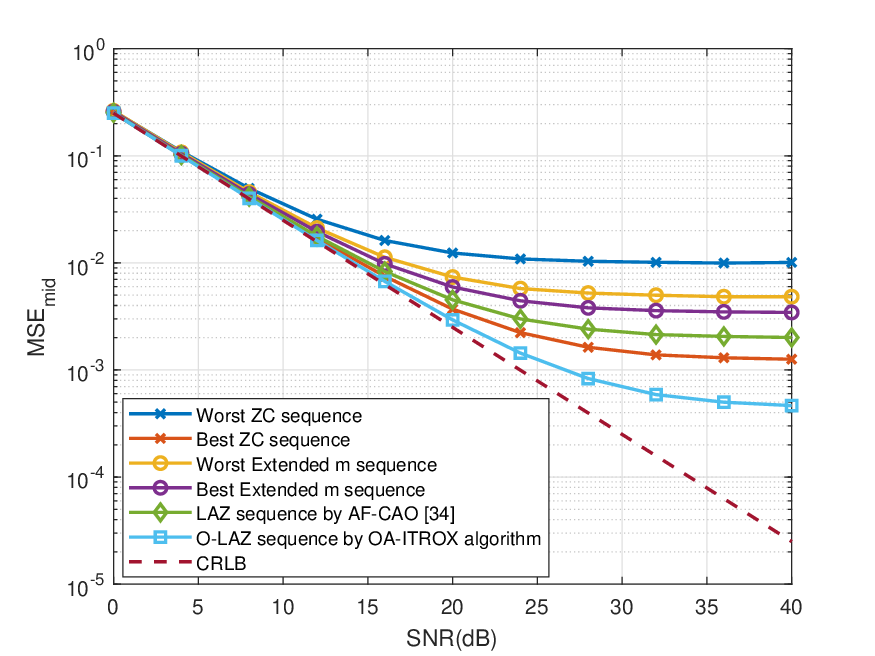}
			\caption{The MSE of the midpoint instant from six pilots.}\label{fig_MSE_mid}
		\end{figure}

%		Table \ref{table_CR} shows the criterion of pilot design for the six sequences mentioned above.
		Fig. \ref{fig_MSE_mid} shows the MSE of the midpoint instant from five pilots.
		The dashed line ``CRLB" in Fig. \ref{fig_MSE_mid} represents the Cramer-Rao Lower Bound (CRLB) under time-invariant conditions, where $\mathrm{CRLB} = \frac{N_\mathrm{CP}}{N}  \sigma^2$ \cite{spasojevic2001complementary, gong2013large,gu2023asymptotically}.
		From Fig. \ref{fig_MSE_mid}, the O-LAZ sequence outperforms the other four sequences significantly with respect to channel estimation performance.
		{From Fig. \ref{fig_MSE_mid}, the proposed O-LAZ sequence exhibits about a 6 dB gain compared to the best ZC sequence.}
		Moreover, the estimation performance of the O-LAZ sequence is very close to the CRLB {when the SNR is below 20 dB.}
		This indicates that the O-LAZ sequence can effectively suppress the ICI interference caused by DSC, thereby achieving channel estimation performance close to that of a time-invariant channel.
%		\begin{table}[ht]
%			\centering
%			\caption{The criterion of pilot design.}\label{table_CR}
%			\begin{tabular}{|c|c|c|c|}
%				\hline
%				% after \\: \hline or \cline{col1-col2} \cline{col3-col4} ...
%				Sequence & $\mathrm{CR}_1$ & $\mathrm{CR}_2$ & $\mathrm{CR}_3$   \\ \hline
%				The worst ZC & 0 & 0.4951 & 275.0104 \\
%				The best ZC & 0 & 0.0537 & 31.1027 \\
%				The worst extended $m$ & 127.7498 & 0.1234 & 809.6574 \\
%				The best extended $m$ & 100.2397 & 0.1469 & 455.8847 \\
%				Traditional LAZ sequence & 17.9414 & 0.0565 & 100.2922 \\
%				\red{O-LAZ sequence} & 6.7876 & 0.0768 & 8.3986 \\
%				\hline
%			\end{tabular}
%		\end{table}
		
		The second simulation experiment considers channel estimation performance and bit error rate (BER) performance for multiple OFDM symbols.
		Here, we adopt the transmission frame structure as shown in Fig. \ref{fig_frame} and use DPSS interpolation to get the channel state information of data symbols. The detailed process of DPSS interpolation is provided in Appendix II. We assume a total of {$M_P = 8$ pilot symbols, with each pair of pilot OFDM symbols containing $M_D=1$} data OFDM symbols.
		All other simulation parameters remain the same as in the previous experiment.
%		In practical engineering scenarios, channel estimation for all OFDM symbols is typically accomplished in two steps.
%		In the first step, channel estimation at the midpoint of pilot OFDM symbols is obtained using pilot symbols.
%		In the second step, channel estimation for all symbols can be obtained using interpolation based on DPSS methods \cite{zemen2003time, zemen2005time}.
		In this experiment, the parameters for generating DPSS interpolation are as follows: The number of DPSS is {8, and the time half bandwidth is 2.02.}
		As described in Subsection II.C, there are a total of $M=M_P+M_D(M_P-1)$ OFDM symbols.
		Hence, the time-domain channel matrix $\mathbf{h}$ corresponding to the OFDM frame with CPs should be an $L\times (N+L)M$ matrix.
		And the $\mathrm{MSE}_\mathrm{frame}$ of the total OFDM frame is defined as follows:
		$$\mathrm{MSE}_\mathrm{frame}=\frac{E\left(\|\mathbf{h}-\hat{\mathbf{h}}\|_F^2\right)}{(N+L)M}.$$
		
\begin{figure}[ht]
			\centering
			\includegraphics[width=8cm]{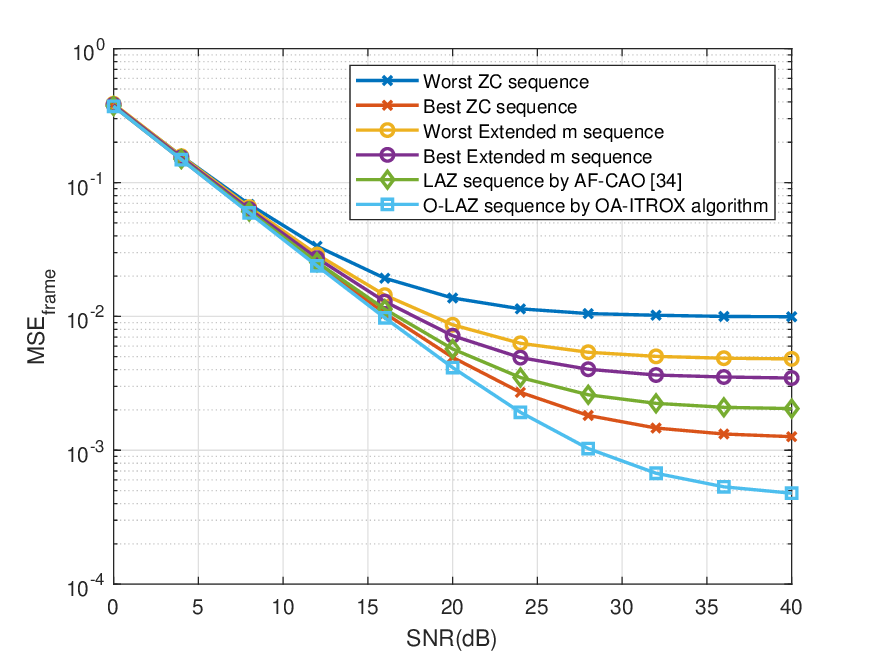}
			\caption{The MSE of the OFDM frame from six pilots.}\label{fig_MSE_all}
		\end{figure}

		Fig. \ref{fig_MSE_all} shows the MSE of the OFDM frame from three pilots.
		Although the DPSS interpolation process introduces new errors, the overall estimation performance remains quite satisfactory.
		The proposed O-LAZ sequence outperforms the best ZC sequence, extended $m$ sequence, and traditional LAZ sequence, significantly in terms of channel estimation MSE.
		{From Fig. \ref{fig_MSE_all}, the O-LAZ sequence exhibits about a 3 dB gain compared to the best ZC sequence.}
		
\begin{figure}[ht]
			\centering
			\includegraphics[width=8cm]{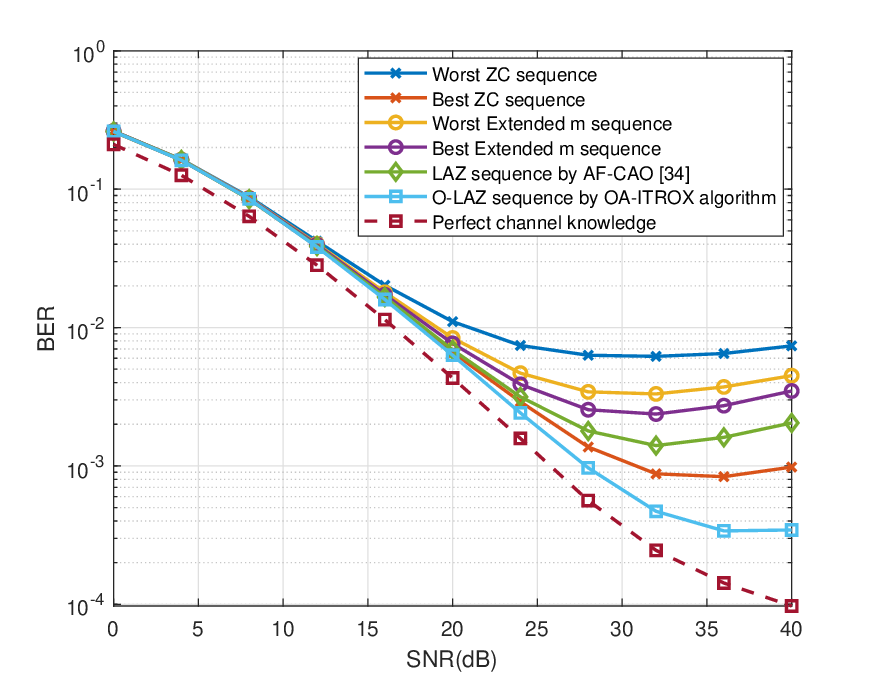}
			\caption{The BER of the data symbols using different pilot sequences.}\label{fig_BER}
		\end{figure}

		In Fig. \ref{fig_BER}, we employed quadrature phase shift keying (QPSK) as the modulation scheme for data symbols. Fig. \ref{fig_BER} shows the BER performance of data symbols with various pilot sequences. The dashed lines represent the BER at the receiver with perfect knowledge of the channel state information (CSI). It is evident from Fig. \ref{fig_BER} that the O-LAZ sequences outperform the best ZC sequence, extended $m$ sequence, and traditional LAZ sequence, in terms of the BER performance. Moreover, the performance of the O-LAZ sequences closely approaches that of the dashed lines.
		
		\section{Conclusion}
		In this paper, we have introduced O-AF and shown that sequences with low O-AF magnitude are suitable for performing channel estimation under DSC, without guard subcarriers, when the channel is approximated through Jakes' model of the Rayleigh fading channel. We have then proposed O-LAZ sequences using a modified ITROX algorithm. Finally, through numerical experiments we have demonstrated the robustness of the algorithm and efficiency of the designed O-LAZ sequences for efficient channel estimation under DSC as compared to the traditional ZC and $m$ sequences.

  \section*{Acknowledgment}
The authors are very grateful to the associate editor, Prof. Sajid Ahmed, and the anonymous reviewers for their valuable comments and suggestions that greatly improved the presentation and quality of this article.
\bibliographystyle{IEEEtran}%{ieeetr}
\bibliography{Refsnew}

\section*{Appendix I}
Note that (\ref{equation_ICI2}) has two random variables $g_q^{(l)}$ and $\omega_q^{(l)}$. Out of the two random variables, the randomness of $\omega_q^{(l)}$ needs to be eliminated to ensure that the resultant sequence efficiently estimates the channel under any channel parameters.
%From (\ref{equation_ICI2}), we can obtain it is not affected by the random variables $g_q^{(l)}$. However, the randomness of (\ref{equation_ICI2}) from $\omega_q^{(l)}$ also needs to be eliminated (a good sequence should be able to work under any channel parameters, its performance needs to be independent of the randomness of the channel).
		Hence, $\mathrm{CR}_2$ is modified from (\ref{equation_ICI2}) to (\ref{equation_ICI3}).

			\begin{equation}\label{equation_ICI3}
				\arg \min_\mathbf{a}
				\int_{-F_r}^{F_r}\cdots\int_{-F_r}^{F_r} \left\| (\mathbf{A}_T^\mathrm{H}\mathbf{A}_T)^{-1} \mathbf{A}_T^\mathrm{H}(\mathbf{A}_T \odot \mathbf{\Omega}_q) -\mathrm{diag}(\mathbf{\Gamma}_q) \right\|^2_F \mathrm{d}\omega_q^{(0)} \cdots \mathrm{d}\omega_q^{(L-1)}.
			\end{equation}

For efficient channel estimation over DSC, pilot sequences need to satisfy both (\ref{Opt_criterion1}) and (\ref{equation_ICI3}). However, satisfying both (\ref{Opt_criterion1}) and (\ref{equation_ICI3}) can be a challenging optimization problem.
			Hence, we re-design (\ref{Opt_criterion1}) and (\ref{equation_ICI3}) using the ISL of the O-AF of sequence $\mathbf{a}$ and combine $\mathrm{CR}_1$ and $\mathrm{CR}_2$ to a single optimization problem.
			
			Note that minimizing (\ref{Opt_criterion1}) is equivalent to minimizing $|\widetilde{AF}_{\mathbf{a}}(\tau,0)|$. Therefore, if we minimize the ISL, $\mathrm{CR}_1$ will also be satisfied.
			Next, if $\mathbf{A}_T^H\mathbf{A}_T=N\mathbf{I}$, then (\ref{equation_ICI3}) can be simplified to (\ref{equation_CR2_2}).

			\begin{equation}\label{equation_CR2_2}
				\arg \min_\mathbf{a}
				\int_{-F_r}^{F_r}\cdots\int_{-F_r}^{F_r} \left\|\frac{1}{N} \mathbf{A}_T^H(\mathbf{A}_T \odot \mathbf{\Omega}_q) - \mathrm{diag}(\mathbf{\Gamma}_q) \right\|^2_F \mathrm{d}\omega_q^{(0)} \cdots \mathrm{d}\omega_q^{(L-1)}.
			\end{equation}

For the ease of expression, we denote the integrand function in (\ref{equation_CR2_2}) as:
\begin{equation}\label{equation_eta1}
  \eta_1(\omega_q^{(0)}, \cdots, \omega_q^{(L-1)})=\left\|\frac{1}{N} \mathbf{A}_T^H(\mathbf{A}_T \odot \mathbf{\Omega}_q) - \mathrm{diag}(\mathbf{\Gamma}_q) \right\|^2_F.
\end{equation}
And (\ref{equation_CR2_2}) can be rewritten as:
\begin{equation}\label{equation_CR2_3}
				\arg \min_\mathbf{a}
				\int_{-F_r}^{F_r}\cdots\int_{-F_r}^{F_r} \eta_1(\omega_q^{(0)}, \cdots, \omega_q^{(L-1)}) \mathrm{d}\omega_q^{(0)} \cdots \mathrm{d}\omega_q^{(L-1)}.
			\end{equation}

Expanding (\ref{equation_eta1}), we obtain (\ref{equation_eta1_2}).
		
		\begin{figure*}[ht]
\begin{multline}\label{equation_eta1_2}
  \eta_1(\omega_q^{(0)}, \cdots, \omega_q^{(L-1)})=  \\
  \left\|\frac{1}{N}
				\left[
				\begin{array}{cccc}
					\sum_{k=0}^{N-1}a_k^*a_k\zeta_N^{\omega_q^{(0)}k} & \sum_{k=0}^{N-1}a_k^*a_{k-1}\zeta_N^{\omega_q^{(1)}k} & \cdots & \sum_{k=0}^{N-1}a_k^*a_{k-L+1}\zeta_N^{\omega_q^{(L-1)}k} \\
					\sum_{k=0}^{N-1}a_{k-1}^*a_k\zeta_N^{\omega_q^{(0)}k} & \sum_{k=0}^{N-1}a_{k-1}^*a_{k-1}\zeta_N^{\omega_q^{(1)}k} & \cdots & \sum_{k=0}^{N-1}a_{k-1}^*a_{k-L+1}\zeta_N^{\omega_q^{(L-1)}k} \\
					\vdots & \vdots &   & \vdots \\
					\sum_{k=0}^{N-1}a_{k-L+1}^*a_k\zeta_N^{\omega_q^{(0)}k} & \sum_{k=0}^{N-1}a_{k-L+1}^*a_{k-1}\zeta_N^{\omega_q^{(1)}k} & \cdots & \sum_{k=0}^{N-1}a_{k-L+1}^*a_{k-L+1}\zeta_N^{\omega_q^{(L-1)}k} \\
				\end{array}
				\right]
				-\mathrm{diag}(\mathbf{\Gamma}_q) \right\|^2_F.
\end{multline}
\end{figure*}

According to the definition of O-AF, (\ref{equation_eta1_2}) can be approximated by (\ref{equation_eta2}).
\begin{figure*}[ht]
\begin{multline}\label{equation_eta2}
  \eta_2(\omega_q^{(0)}, \cdots, \omega_q^{(L-1)})=
  \left\|\frac{1}{N}
				\left[
				\begin{array}{cccc}
					\widetilde{AF}_{\mathbf{a}}(0,\omega_q^{(0)}) & \widetilde{AF}_{\mathbf{a}}(-1,\omega_q^{(1)}) & \cdots & \widetilde{AF}_{\mathbf{a}}(1-L,\omega_q^{(L-1)}) \\
					\widetilde{AF}_{\mathbf{a}}(1,\omega_q^{(0)}) & \widetilde{AF}_{\mathbf{a}}(0,\omega_q^{(1)}) & \cdots & \widetilde{AF}_{\mathbf{a}}(2-L,\omega_q^{(L-1)}) \\
					\vdots & \vdots &   & \vdots \\
					\widetilde{AF}_{\mathbf{a}}(L-1,\omega_q^{(0)}) & \widetilde{AF}_{\mathbf{a}}(L-2,\omega_q^{(1)}) & \cdots & \widetilde{AF}_{\mathbf{a}}(0,\omega_q^{(L-1)}) \\
				\end{array}
				\right]
				-\mathrm{diag}(\mathbf{\Gamma}_q) \right\|^2_F.
\end{multline}
\end{figure*}

Hence, (\ref{equation_CR2_3}) can be approximated as follows:
\begin{equation}\label{equation_CR2_4}
	\arg \min_\mathbf{a} \int_{-F_r}^{F_r}\cdots\int_{-F_r}^{F_r} \eta_2(\omega_q^{(0)}, \cdots, \omega_q^{(L-1)}) \mathrm{d}\omega_q^{(0)} \cdots \mathrm{d}\omega_q^{(L-1)}.
\end{equation}

Since the values of $\omega_q^{(l)},$ for each $l=0,\ldots,L-1$ lies within $[-F_r,F_r]$, the minimization problem of the multiple integral in (\ref{equation_CR2_4}) can be approximated as the minimization problem of a single integral \cite{Wazwaz2011}, as follows:
		\begin{equation}\label{equation_CR2_5}
			\arg \min_\mathbf{a} \int_{-F_r}^{F_r}\eta_3(f) \mathrm{d}f,
		\end{equation}
		where the integrand $\eta_3(f)$ is given by (\ref{equation_eta3}).

\begin{figure*}[ht]
			\begin{equation}\label{equation_eta3}
				\eta_3(f)=\left\|\frac{1}{N} \left(
				\left[
				\begin{array}{cccc}
					\widetilde{AF}_{\mathbf{a}}(0,f) & \widetilde{AF}_{\mathbf{a}}(-1,f) & \cdots & \widetilde{AF}_{\mathbf{a}}(1-L,f) \\
					\widetilde{AF}_{\mathbf{a}}(1,f) & \widetilde{AF}_{\mathbf{a}}(0,f) & \cdots & \widetilde{AF}_{\mathbf{a}}(2-L,f) \\
					\vdots & \vdots &   & \vdots \\
					\widetilde{AF}_{\mathbf{a}}(L-1,f) & \widetilde{AF}_{\mathbf{a}}(L-2,f) & \cdots & \widetilde{AF}_{\mathbf{a}}(0,f) \\
				\end{array}
				\right]
				-\sum_{k=0}^{N-1}\zeta_N^{fn}\mathbf{I}_L\right) \right\|^2_F.
			\end{equation}
		\end{figure*}
		
		The problem of minimizing (\ref{equation_CR2_5}) can be approximately replaced by the following minimization problem:
			\begin{equation}\label{equation_CR2_6}
				\arg \min_\mathbf{a} \int_{-F_r}^{F_r}
				\frac{1}{N}\left(
				\sum_{\substack{\tau=-Z\\ \tau\neq0}}^{Z} |\widetilde{AF}_{\mathbf{a}}(\tau,f)|^2
				\right)\mathrm{d}f.
			\end{equation}

By exchanging the order of integration and summation, we have
\begin{equation}
			\arg \min_\mathbf{a} \frac{1}{N}
			\sum_{\substack{\tau=-Z\\ \tau\neq0}}^{Z} \int_{-F_r}^{F_r} |\widetilde{AF}_{\mathbf{a}}(\tau,f)|^2\mathrm{d} f.
		\end{equation}

\section*{Appendix II}
We adopt the transmission frame structure as shown in Fig. 2.
It is assumed that the midpoint instant channel response of the pilot symbols is obtained by (\ref{equation_LS}).
Let $\mathbf{h}_{\text{frame}}$ denotes the channel response of the $M$ OFDM symbols in an OFDM frame, based on the DPSS representation principle, we have
\begin{equation}\label{}
  \mathbf{h}_{\mathrm{frame}}^{(l)}=\sum_{k=0}^{B-1} g_k^{(l)} \mathbf{b}_k +\mathbf{e}^{(l)},
\end{equation}
where $\mathbf{h}_{\mathrm{frame}}^{(l)}$ is the channel response corresponding to the $l$-th path of $\mathbf{h}_{\mathrm{frame}}$, $B$ is the number of DPSSs, $\mathbf{b}_k$ is the $k$-th DPSS, $\mathbf{e}^{(l)}$ represents the error in the channel reconstruction using DPSS. Note that $\mathbf{h}_{\mathrm{frame}}^{(l)}$ contains the channel responses corresponding to $M$ OFDM symbols, and among these, we only know the values at the midpoints of $M_P$ pilot symbols, i.e., $[h_{\mathrm{frame}}^{(l)}[p_0], h_{\mathrm{frame}}^{(l)}[p_1], \ldots, h_{\mathrm{frame}}^{(l)}[p_{M_P-1}]]^\mathrm{T}$.
To solve for all the coefficients $g_k^{(l)}$, it is generally required that $M_P \geq B$.
Without loss of generality, we assume that $M_P = B$.
Then, all the coefficients $g_k^{(l)}$ can be obtained from (\ref{equation_DPSS}), and $\mathbf{h}_{\mathrm{frame}}$ can also be calculated once $g_k^{(l)}$ is obtained, through routine calculation.

\begin{equation}\label{equation_DPSS}
\left[
  \begin{array}{c}
    g_0^{(l)} \\
    g_1^{(l)} \\
    \vdots \\
    g_{B-1}^{(l)} \\
  \end{array}
\right]
=
\left[
  \begin{array}{cccc}
    b_0[p_0] & b_0[p_1] & \cdots & b_0[p_{M_P-1}] \\
    b_1[p_0] & b_1[p_1] & \cdots & b_1[p_{M_P-1}] \\
    \vdots & \vdots &  & \vdots \\
    b_{B-1}[p_0] & b_{B-1}[p_1] & \cdots & b_{B-1}[p_{M_P-1}] \\
  \end{array}
\right]^{-1}
\left[
  \begin{array}{c}
    h_{\mathrm{frame}}^{(l)}[p_0] \\
    h_{\mathrm{frame}}^{(l)}[p_1] \\
    \vdots \\
    h_{\mathrm{frame}}^{(l)}[p_{M_P-1}] \\
  \end{array}
\right].
\end{equation}

		%	\section*{Acknowledgments}
		%	This should be a simple paragraph before the References to thank those individuals and institutions who have supported your work on this article.

	\end{document}